\documentstyle[amssymb,aps]{revtex}

\newtheorem{theorem}{Theorem}

\newtheorem{lemma}[theorem]{Lemma}

\begin{document}
\def\rel{\sqrt{\mu^2-\lambda_n^2}}
\def\beq{\begin{eqnarray}}
\def\eeq{\end{eqnarray}}
\def\bea{\begin{array}}
\def\eea{\end{array}}
\def\gaa{\gamma_a^T}
\newcommand{\nats}{\mbox{${\rm I\!N }$}}
\def\gab{\gamma_b^T}
\def\gm{\gamma_m}
\def\cz {{\cal Z}_n}
\def\czp{{\cal Z}_n^\pm}
\def\ph{\hat\psi}
\def\phc{\hat\psi ^*}
\def\nn{\nonumber}
\def\ep{\epsilon}
\def\operatorname#1{{\rm#1\,}}
\def\text#1{{\hbox{#1}}}
\def\qedbox{\hbox{$\rlap{$\sqcap$}\sqcup$}}
\def\qed{\nobreak\hfill\penalty250 \hbox{}\nobreak\hfill\qedbox}
\def\proof{\medbreak\noindent{\bf Proof:} \rm }
\def\trace{\operatorname{Tr}}
\def\Trace\{#1\}{\operatorname{Tr}\{#1\}}
\def\dzero{\frac 1 {32} (1-\frac{\beta(m)} {m-2} )}
\def\dtwo{\frac 1 {16}  ( 5-2m+\frac{7-8m+2m^2}{m-2} \beta (m)) }
\def\dthree{\frac 1 {32 (m-1)}(2m-3 -\frac{2m^2-6m+5}{m-2}\beta (m) )}
\def\dfive{\frac 1 {16 (m-1)}(1+\frac{3-2m}{m-2} \beta (m)) }
\font\small=cmr7\def\nmonth{\ifcase\month\ \or January\or
February\or March\or April\or May\or June\or July\or August\or 
September\or October\or November\else December\fi} 
\def\DATE{\nmonth\ \number\day\ \number\year}

\title{Heat asymptotics with spectral boundary conditions II}
\author{Peter Gilkey\thanks{Research Partially supported by the NSF (USA)
     and MPI (Leipzig)}}
\address{Mathematics Department, University of Oregon, Eugene Or 97403 USA
    \newline\phantom{ab}\quad{\tt http://darkwing.uoregon.edu/$\sim$gilkey}
   email:gilkey@darkwing.uoregon.edu}
\author{Klaus Kirsten\thanks{Research Partially supported by EPSRC under 
Grant No GR/M45726}}
\address{Department of Physics and Astronomy, The University of
     Manchester, Oxford Road, Manchester UK M13 9PL UK
     \newline email: klaus@a13.ph.man.ac.uk}
\maketitle

\begin{abstract}{\bf ABSTRACT:} \rm Let $P$ be an operator of Dirac type on a
compact
   Riemannian manifold with smooth boundary. We impose
   spectral boundary conditions and study the asymptotics of the
   heat trace of the associated operator of Laplace type.\\
   Subject Code: Primary 58G25; PACS numbers: 1100, 0230, 0462.
\end{abstract}

\bigbreak
We recall the notational conventions established in \cite{DGK}.
Let $M$ be a compact $m$-dimensional Riemannian manifold with smooth
boundary $\partial M$. We suppose given unitary vector bundles $E_i$ over $M$ and an elliptic
complex
\begin{equation} P:C^\infty(E_1)\rightarrow C^\infty(E_2).\label{EQNu}\end{equation}
We assume that (\ref{EQNu}) defines an elliptic complex of Dirac type. We impose spectral
boundary conditions ${\mathcal B}$; Atiyah, Patodi, and Singer \cite{APS} showed that an elliptic
complex of Dirac type need not admit local boundary conditions.

Apart from the mathematical interest, spectral boundary conditions are of 
relevance in one-loop quantum cosmology and supergravity 
(see e.g. \cite{giam1,giam2}). Furthermore, they are consistent with a 
non-zero index and have been intensively discussed in the context of 
fermion number fractionization \cite{frac1,frac2}. 

Let
$P_{\mathcal B}$ and
$D_{\mathcal B}:=(P_{\mathcal B})^*P_{\mathcal B}$ be the associated realizations. Let $F\in
C^\infty(E_1)$ be an auxiliary function used for localization. Results of Grubb and Seeley
\cite{GSa,GSb,GSc} show that there is an asymptotic series as $t\downarrow0$ of the
form:
\begin{equation}\operatorname{Tr}_{L^2}\{Fe^{-tD_{{\mathcal B}}}\}\sim
 \sum_{0\le k\le m-1}a_k(F,D,{\mathcal B})t^{(k-m)/2}+O(t^{-1/8}).
 \label{EQNb}\end{equation}
(There is in fact a complete asymptotic series with log terms, but we shall only be interested in
the first few terms in the series). The coefficients $a_k$ in equation (\ref{EQNb}) are locally
computable. We determined the coefficients $a_0$, $a_1$, and $a_2$ previously \cite{DGK}; these
results are summarized in Theorem
\ref{THM1} below. In this paper, we determine the coefficient $a_3$. We shall assume henceforth
that $m\ge4$ so that the series in equation (\ref{EQNb}) gives this term.

We shall express the coefficients $a_k$ invariantly in terms of the following data. Let $\gamma$ be
the leading symbol of the operator $P$. Since the elliptic complex is of Dirac type,
$\gamma+\gamma^*$ defines a unitary Clifford module structure on $E_1\oplus E_2$. Let
$\nabla=\nabla_1\oplus\nabla_2$ be a compatible unitary connection; this means that
\begin{equation}\nabla(\gamma+\gamma^*)=0 \text{ and }
  (\nabla s,\tilde s)+(s,\nabla\tilde s)=d(s,\tilde s).\label{EQNa}\end{equation}
Such connections always exist \cite{BGb} but are not unique. If
$y=(y^1,...,y^{m-1})$ are local coordinates on $\partial M$, let $x=(y,x^m)$ be
local coordinates on the collar where $x^m$ is the geodesic distance to the
boundary; the curves $y\rightarrow (y,t)$ are unit speed geodesics perpendicular to $\partial M$.
Let $\partial_\mu:=\frac{\partial}{\partial x^\mu}$; $\partial_m$ is the {\bf inward} geodesic normal
vector field on the collar. Let $\nabla_\mu$ be covariant differentiation with respect to
$\partial_\mu$. Decompose
$$P=\gamma^\mu\nabla_\mu+\psi$$ 
where we adopt the Einstein convention and sum over repeated indices. Here $\psi$ is a $0^{th}$
order operator; the structures
$\gamma$,
$\nabla$, and
$\psi$ can depend on the normal variable. Since $P$ is of Dirac type, we have the Clifford
commutation relations:
\begin{equation}(\gamma^\mu)^*\gamma^\nu+(\gamma^\nu)^*\gamma^\mu=2g^{\mu\nu}.
   \label{Clifrel}\end{equation}

Near the boundary and relative
to a local frame which is parallel along the normal geodesic rays, we have
$\nabla_m=\partial_m$. We freeze the coefficients and set
$x^m=0$ to define a tangential operator
$$B_{}(y):=\gamma^m(y,0)^{-1}\{\textstyle
\sum_{\alpha<m}\gamma^\alpha(y,0)\nabla_\alpha+\psi(y,0)\}
\text{ on }C^\infty(E_1|_{\partial M}).$$
Let
$\Theta$ be an auxiliary self-adjoint endomorphism of $E_1|_{\partial M}$.
We take the adjoint of $B_{}$ with respect to the structures on the boundary to define
a self-adjoint tangential operator of Dirac type on $C^\infty(E_1|_{\partial M})$:
$$A:= {B_{}+B_{}^*\over2}+\Theta.$$  
The boundary operator ${\mathcal B}$ whose vanishing defines spectral boundary conditions is
orthogonal projection on the span of the eigenspaces for the non-negative spectrum of
$A$. Replacing the words ``non-negative'' by ``positive'' would not change the local invariants
$a_n$.

We shall let Roman
indices $i$ and $j$  range from $1$ to $m$ and index a local orthonormal frame for the tangent
bundle of $M$; Greek indices will index a local coordinate frame.  Near the boundary, we
choose the frame so that $e_m$ is the inward unit geodesic normal vector; we
let indices $a$ and $b$ range from $1$ through $m-1$ and index the
corresponding frame for the tangent bundle of the boundary. We adopt the
Einstein convention and sum over repeated indices. We let `;' denote
multiple covariant differentiation of the tensors involved.
Let $\Gamma$ be the Christoffel symbols of the Levi-Civita connection on
$M$. There is a canonical connection ${}^D\nabla$ on the bundle $E_1$ and there is a canonical
endomorphism $E$ of the bundle $E_1$ so that $D=-(\Trace\{{}^D\nabla^2\}+E)$; see \cite{Gb} for
details. Note that ${}^D\nabla$ is not in general a compatible connection. Let
$\omega$ be the connection
$1$ form of ${}^D\nabla$. We have the following equations of structure:
\begin{eqnarray}
&&D=-(g^{\mu\nu}\partial_\mu\partial_\nu+a^\mu\partial_\mu+b)
   =-(\Trace\{{}^D\nabla^2\}+E),\nonumber\\
&&\omega_\delta:= \textstyle{\frac 1 2}g_{\nu\delta}(a^\nu+g^{\mu\sigma}
    \Gamma_{\mu\sigma}{}^\nu),\text{ and }\label{EQNc}\\
&&E:=b-g^{\nu\mu}(\partial_\nu\omega_\mu
   +\omega_\nu\omega_\mu-\omega_\sigma\Gamma_{\nu\mu}{}^\sigma).\nonumber
   \end{eqnarray}
Decompose $P=\gamma_i\nabla_i+\psi$. Let $R_{ijkl}$ be the Riemann curvature tensor. Let
\begin{eqnarray}
&&\hat\psi:=\gamma_m^{-1}\psi,\ \tau:=R_{ijji},\ \rho_{ij}=R_{ikkj},\\
  &&\beta(m):=\Gamma(\textstyle\frac m2)\Gamma(\textstyle\frac12)^{-1}\Gamma(\frac{m+1}2)^{-1}.
    \label{EQNd}\end{eqnarray}
Let `;' and `:' denote multiple covariant differentiation with respect to the background connection $\nabla$ and
the Levi-Civita connections on $M$ and $\partial M$ respectively. Let $L_{ab}:=\Gamma_{abm}$ be the second fundamental
form. The following is the main result of this paper.\def\bork{\phantom{\vrule height .5true cm\qquad\qquad\qquad}}
\begin{theorem}\label{THM1} We have\begin{enumerate}
\item $a_0(F,D,{\mathcal B}) =  (4\pi )^{-m/2} \int_{M}\Trace\{F\}$.
\item 
      $a_1(F,D,{\mathcal B}) = (4\pi)^{-(m-1)/2}\frac14(\beta(m)-1)\int_{\partial M}\Trace\{F\}$.
\item $a_2(F,D,{\mathcal B})
     =(4\pi)^{-m/2}\int_M\frac{1}{6}\Trace
         \{F(\tau+6E)\}
   +(4\pi)^{-m/2}\int_{\partial M}\Trace\{
     \frac{1}{2}[\hat\psi+\hat\psi^*]F+\frac13(1-
     \frac34\pi \beta(m))L_{aa} F$\newline
     \bork $-\frac{m-1}{2(m-2)}(1-
        \frac12 \pi \beta(m))F_{;m}\}.$
\smallskip\item $a_3(F,D,{\mathcal B})=(4\pi)^{-(m-1)/2}\textstyle\int_{\partial M}F\Trace\{
       \dzero (\hat\psi \hat\psi+\hat\psi^* \hat\psi^* )$
  $     +\dtwo\hat\psi\hat\psi^* $\newline\bork
   $   +\dthree(\gamma_a^T\hat\psi\gamma_a^T\hat\psi
       +\gamma_a^T\hat\psi^*\gamma_a^T\hat\psi^*)$
      $+\dfive
       \gamma_a^T\hat\psi\gamma_a^T\hat\psi^*$\newline\bork$
    -\frac1{48}(\frac{m-1}{m-2}\beta(m)-1)\tau$
    $+\frac1{48}(1- \frac{4m-10}{m-2}\beta(m))\rho_{mm}$
   $+\frac 1 {48 (m+1)}(\frac{17+5m} 4 +\frac{23-2m-4m^2}{m-2} \beta (m)
   ) L_{ab}L_{ab} $\newline\bork
 $  +\frac 1 {48 (m^2-1)}(-\frac{17+7m^2}{8} + \frac{4m^3-11m^2+5m-1}
         {m-2} \beta (m)) L_{aa} L_{bb}$
$+\frac 1 {8(m-2)} \beta (m)(\Theta \Theta +\frac 1 {m-1}
     \gamma_a^T\Theta \gamma_a^T\Theta)\}$\newline\bork
    $+\frac 1 {8(m-3)} ( \frac{5m-7} 8 -\frac{5m-9} 3 
           \beta (m)) L_{aa}F_{;m}\Trace\{I\}
    +\frac{m-1}{16(m-3)}(2\beta(m)-1)F_{;mm}\Trace\{I\}$.
\end{enumerate}\end{theorem}

We refer to
\cite{DGK} for the proof of assertions (1)-(3); 
the remainder of this article is devoted to the proof of assertion (4). 
We begin by giving a general recipe for the invariant $a_3$.
The coefficients $a_{2k}$ involve both interior and boundary integrals. The coefficients
$a_{2k+1}$ only involve boundary integrals.
One can use dimensional analysis to see that the boundary integrand for $a_3$ can be expressed in
terms of local invariants which are homogeneous of order
$2$ in the jets of the total symbols. Since $P=\gamma^\nu\nabla_\nu+\psi$, we must consider
invariant expressions determined by the jets of $\gamma$, $\nabla$, and $\psi$. Instead of using the curvature
$\Omega_{ij}$ of the background connection $\nabla$ as one of our basic invariants, we shall instead
use the tensor 
\begin{equation}W_{ij}:=\textstyle\Omega_{ij}-\frac 1 4 R_{ijkl} \gamma_k^*
\gamma_l.\label{EQNe}\end{equation}
Since the background connection $\nabla$ is assumed to be
compatible, we have
$[\gamma,W]=0$; see \cite{BGb} for details. Since $\nabla\gamma=0$, the covariant derivatives of
$\gamma$ do not enter. We define:
\begin{equation}\gamma_a^T:=\gamma_m^{-1}\gamma_a\text{ for }1\le a\le
m-1.\label{EQNf}\end{equation}
Equation (\ref{Clifrel}) implies that $\gamma^T$ is a unitary Clifford module structure, i.e. that
we have the relations:
\begin{equation}\gamma_a^T\gamma_b^T+\gamma_b^T\gamma_a^T=-2\delta_{ab}\text{ and
}(\gamma_a^T)^*=-\gamma_a^T.\label{ClifrelT}\end{equation}
The Grubb-Seeley calculus controls the pure
$\gamma$ terms and after a bit of work using the Weyl calculus, one can show the following Lemma
holds where we adopt the notation of equations (\ref{EQNc}-\ref{EQNf})
\begin{lemma}\label{LEM1} There exist universal constants $d_i=d_i(m)$ for $0\le i\le 20$ and
$e_i=e_i(m)$ for $0\le i\le 8$ so that 
$a_3(F,D,{\mathcal B})=(4\pi)^{-(m-1)/2}\int_{\partial M}\Trace\{a_3(F,D,{\mathcal B})(y)\}dy$ where:
\begin{eqnarray*} 
   a_3 (F,D,{\mathcal B}, y)&=& F ( d_0 [\ph \ph + \phc \phc ] + 
                d_1 [\ph \ph - \phc \phc ] +d_2 \phc \ph
+d_3 [\gaa \ph \gaa \ph +\gaa \phc \gaa \phc ]\\& & + 
   d_4 [\gaa \ph \gaa \ph -\gaa \phc \gaa \phc]
  +d_5 \gaa \phc \gaa \ph +d_6 [\ph _{;m} + \phc _{;m}] 
   +d_7 [\ph _{;m} -\phc _{;m}]\\&&
   +d_8 [\gaa \ph _{:a} +\gaa \phc _{:a} ] +d_9 [ \gaa \ph _{:a} 
       -\gaa \phc _{:a} ]
   +d_{10} L_{aa}[\ph + \phc ]
   +d_{11}L_{aa} [ \ph - \phc]
  \\&& +d_{12} \tau +d_{13} \rho_{mm} 
   +d_{14} W_{ab} \gaa \gab \nn
   +d_{15} W_{am} \gaa +d_{16} L_{ab} L_{ab} +d_{17} L_{aa} L_{aa})  \\
& &+F_{;m} (d_{18} [\ph +\phc] +d_{19} [\ph -\phc ] +d_{20} L_{aa} ) 
   +d_{21} F_{;mm} \nn\\
& &+F (e_0 \Theta \Theta +e_1 \gaa \Theta \gaa \Theta 
        +e_2 \gaa \Theta_{:a} +e_3 L_{aa} \Theta 
   +e_4 \Theta [\ph +\phc ] +e_5 \Theta [\ph -\phc ] \\
& &+e_6 \gaa \Theta \gaa [\ph +\phc ] +e_7 \gaa \Theta \gaa [\ph -\phc ] ) 
         +e_8 F_{;m} \Theta.
\end{eqnarray*}\end{lemma}

We prove Theorem \ref{THM1} by determining the unknown constants of Lemma \ref{LEM1}.
We first establish some technical results.

\begin{lemma}\label{d9}\ \begin{enumerate}
\smallskip\item We have that $d_9\textstyle\int_{\partial
M}F\Trace\{\gamma_a^T(\hat\psi-\hat\psi^*)_{:a}\} =-d_9\textstyle\int_{\partial
M}F_{:a}\Trace\{\gamma_a^T(\hat\psi-\hat\psi^*)\}$.
\smallskip\item The dual boundary condition for the formal adjoint
$P^*$ is projection on the non-negative spectrum of the operator $A_2:=-\gamma_m A_1\gamma_m^{-1}$. Furthermore
$A_2$ is defined by $\Theta_2=-\gamma_m\Theta_1\gamma_m^{-1}+L_{aa}$.
\end{enumerate}
\end{lemma}

\medbreak\noindent{\bf Proof.} We shall derive equation (2.18) of \cite{G} showing $\gamma_{a:a}^T=0$; assertion(1)  then
follows by integration by parts:
\begin{eqnarray*}
\gamma_{a:a}^T&=&\nabla_a\gamma_a^T-\gamma_a^T\nabla_a-\Gamma_{aab}\gamma^T_b
      =-\nabla_a\gamma_m\gamma_a+\gamma_m\gamma_a\nabla_a+\Gamma_{aab}\gamma_m\gamma_b\\
&=&-(\nabla_a\gamma_m-\gamma_m\nabla_a)\gamma_a-\gamma_m(\nabla_a\gamma_a-\gamma_a\nabla_a)
   +\Gamma_{aab}\gamma_m\gamma_b\\
&=&-\gamma_{m;a}\gamma_a-\gamma_m\gamma_{a;a}-\Gamma_{amb}\gamma_b\gamma_a
   -\Gamma_{aai}\gamma_m\gamma_i+\Gamma_{aab}\gamma_m\gamma_b\\
&=&L_{ab}\gamma_b\gamma_a
   -L_{aa}\gamma_m\gamma_m=0.
\end{eqnarray*}
 We compute the
Green's formula to prove assertion (2). We have $\gamma^\nu:E_1\rightarrow E_2$. The operator $\gamma^\nu:E_2\rightarrow E_1$
was defined by $(\phi_1,\gamma^\nu\phi_2)=-(\gamma^\nu\phi_1,\phi_2)$. We compute:
\begin{eqnarray}
&&(P\phi_1,\phi_2)_{L^2}-(\phi_1,P^*\phi_2)_{L^2}=
\textstyle\int_M(\gamma^\nu\nabla_\nu\phi_1,\phi_2)-(\phi_1,\nabla_\nu\gamma^\nu\phi_2)
=\textstyle\int_M\partial_\nu(\gamma^\nu\phi_1,\phi_2)=-\int_{\partial M}(\gamma_m\phi_1,\phi_2).\label{green}
\end{eqnarray}
We introduce the following tangential partial differential operators:
$$B_1:=\gamma_m^{-1}\gamma_a\nabla_a+\gamma_m^{-1}\psi_1,\quad
  A_1:=\textstyle{1\over2}(B_1+B_1^*)+\Theta_1,\text{ and }A_2:=-\gamma_mA_1\gamma_m^{-1}.$$
Let $E(\lambda,A_i):=\{\phi\in C^\infty(E_i|_{\partial M}):A_i\phi=\lambda\phi\}$ be the eigenspaces of $A_i$. Let
\begin{eqnarray*}
&&{\mathcal L}_1^{>}:=\text{Closed span}\{E(\lambda,A_1):\lambda>0\},\quad
  {\mathcal L}_1^\le:=\text{Closed span}\{E(\lambda,A_1):\lambda\le0\},\\
&&{\mathcal L}_2^\ge:=\text{Closed span}\{E(\lambda,A_2):\lambda\ge0\},\quad
  {\mathcal L}_2^<:=\text{Closed span}\{E(\lambda,A_2):\lambda<0\}.
\end{eqnarray*}
We then have orthogonal direct sum decompositions 
$$L^2(E_1|_{\partial M})={\mathcal L}_1^>\oplus{\mathcal L}_1^\le\text{ and }
  L^2(E_2|_{\partial M})={\mathcal L}_2^\ge\oplus{\mathcal L}_2^<.$$
Let ${\mathcal B}_i\phi_i$ be orthogonal projection of $\phi_i|_{\partial M}$ on ${\mathcal
L}_1^>$ and
${\mathcal L}_2^\ge$ respectively. As $\gamma_mA_1=-A_2\gamma_m$, $\gamma_mE(\lambda,A_1)=E(-\lambda,A_2)$. Consequently
we have that
$$\gamma_m{\mathcal L}_1^>={\mathcal L}_2^<\text{ and }
  \gamma_m{\mathcal L}_1^\le={\mathcal L}_2^\ge.$$
Let $\phi_i\in C^\infty(E_i)$. We have $\phi_1\in\text{Domain}(P)$ if
and only if 
${\mathcal B}_1\phi_1=0$ or equivalently if $\phi_1|_{\partial M}\in{\mathcal L}_1^\le$.
We use equation (\ref{green}) to see that the following  
assertions are equivalent:
\medbreak\qquad\qquad(1) $\phi_2\in\text{Domain}(P^*)$.
\smallbreak\qquad\qquad(2) $(\gamma_m\phi_1,\phi_2)_{L^2(\partial M)}=0$ for every $\phi_1\in\text{Domain}(P)$.
\smallbreak\qquad\qquad(3) $\phi_2|_{\partial M}\in\{\gamma_m{\mathcal L}_1^\le\}^\perp=({\mathcal L}_2^\ge)^\perp
   ={\mathcal L}_2^<$
i.e. $\phi_2\in\ker{\mathcal B}_2$.
\medbreak\noindent 
Thus ${\mathcal B}_2$ defines the adjoint
boundary condition. As $\nabla\gamma=0$,
\begin{eqnarray*}
\nabla_a\gamma_m&=&\gamma_m\nabla_a+\Gamma_{amb}\gamma_b=\gamma_m\nabla_a-L_{ab}\gamma_b,\\
B_2:&=&-\gamma_mB_1\gamma_m^{-1}=-\gamma_m\gamma_m^{-1}\gamma_a\nabla_a\gamma_m^{-1}-\psi_1\gamma_m^{-1}\\
    &=&\gamma_m^{-1}\gamma_a\nabla_a-L_{ab}\gamma_a\gamma_b-\psi_1\gamma_m^{-1}
      =\gamma_m^{-1}\gamma_a\nabla_a+L_{aa}-\psi_1\gamma_m^{-1},\\
A_2:&=&-\textstyle{1\over2}\gamma_m(B_1+B_1^*)\gamma_m^{-1}-\gamma_m\Theta_1\gamma_m^{-1}\\
   &=&\textstyle{1\over2}(\gamma_m^{-1}\gamma_a\nabla_a+(\gamma_m^{-1}\gamma_a\nabla_a)^*-\psi_1\gamma_m^{-1}
-\gamma_m\psi_1^*)+L_{aa}-\gamma_m\Theta_1\gamma_m^{-1}.\end{eqnarray*}
On the other hand since $\psi_2=\psi_1^*$ and $\gamma_m^*=\gamma_m^{-1}=-\gamma_m$, we have
\begin{eqnarray*}
A_2&=&\textstyle{1\over2}(\gamma_m^{-1}\gamma_a\nabla_a+(\gamma_m^{-1}\gamma_a\nabla_a)^*
    -\gamma_m\psi_1^*-\psi_1\gamma_m^{-1})+\Theta_2\text{
so }\\
\Theta_2&=&-\gamma_m\Theta_1\gamma_m^{-1}+L_{aa}. \qedbox
\end{eqnarray*}

We use functorial properties of the invariants $a_n$ to establish the following Lemma. Recall that we defined
$$\beta(m):=\Gamma(\textstyle\frac m2)   \Gamma(\textstyle\frac12)^{-1}\Gamma(\frac{m+1}2)^{-1}.$$

\begin{lemma}\label{LEM2}\ \begin{enumerate}
\smallskip\item We have \phantom{2a)} $0=d_1=d_4=d_7=d_8=d_{11}=d_{19}=e_2=e_5=e_7$.
\smallskip\item We have 2a) $0=e_3=e_8$,\newline
    \phantom{We have} 2b) $0=e_0 -(m-1) e_1$, and\newline
    \phantom{We have} 2c) $0=e_4 -(m-1) e_6$.
\smallskip\item We may take $d_{14}=0$ and $d_{15} =0$.
\smallskip\item We have \phantom{2a)} $0=d_6=d_{10}$.
\smallskip\item We have 5a) $0=d_{18}$,\newline
   \phantom{We have}  5b) $0=2(m-1)d_{12}+d_{13}-2d_{16}+2(1-m)d_{17}+(3-m)d_{20}$, and\newline
   \phantom{We have} 5c) $0=2(1-m)d_{12}+(1-m)d_{13}+(3-m)d_{21}$.
\smallskip\item We have 6a) $0=2d_0 +d_2 +(m-3)(2d_3 +d_5)$,\newline
   \phantom{We have} 6b) $0=-2d_0 +d_2 +(m-1)(2d_3 -d_5)$, \newline
   \phantom{We have} 6c) $0=e_4+(m-3)e_6$, and\newline
   \phantom{We have} 6d) $0=d_9$.
\smallskip\item We have \phantom{2a)} $0=-2d_0+d_2-(m-1)(2d_3-d_5)-\frac{m-2}4(\beta(m)-1)$.
\smallskip\item We have $0=\textstyle\frac14(\beta(m)-1)+2d_0+d_2+2(m-1)d_3+(m-1)d_5+e_0+e_1(m-1)-2e_4-2e_6(m-1)$.
\smallskip\item We have 9a) $2d_0+d_2=\frac{m-3}8(
\frac{m-1}{m-2}\beta(m)-1)$\newline
    \phantom{We have} 9b) $2d_3+d_5=-\frac18
   (\frac{m-1}{m-2}\beta(m)-1)$,and\newline
    \phantom{We have} 9c) $d_{12}=-\frac1{48}
     (\frac{m-1}{m-2}\beta(m)-1)$.
\smallskip\item We have 10a) $d_{16}+(m-1)d_{17}=\frac{17-7m}{384}+\frac{4m-11}{48}\beta(m)$,\newline
     \phantom{We have} 10b) $d_{20}=\frac1{8(m-3)}
     (\frac{5m-7}8-\frac{5m-9}3\beta(m))$, and\newline
     \phantom{We have} 10c) $d_{21}=\frac{m-1}{16(m-3)}(-1+2\beta(m))$.
\smallskip\item We have $d_{16}+d_{17}=\frac 1 {16 (m^2-1)} 
        (\frac{m^2+8m-17} 8 - (3m-4) \beta (m)) $.
\end{enumerate}\end{lemma}

\medbreak\noindent{\bf Remark}
We use equations (2c) and (6c) to see $e_4=e_6=0$. 
Equation (6a) is not independent from (9a) and (9b). 
Using (9a) and (9b) in (8), an equation for $e_0$ and $e_1$ follows. Together
with (2b) this determines $e_0$ and $e_1$.
We solve equations (6b), (7), (9a), and (9b) 
to determine $d_0$, $d_2$, $d_3$, and $d_5$. 
Thus we
complete the proof of Theorem
\ref{THM1} (4) by checking that the non-zero coefficients are given
by:\def\bork{\vphantom{$A_{A_{A_A}}^{A^A}$}\ }\smallbreak\noindent
\centerline{\begin{tabular}{| l |l|}\hline
\ $d_0=\dzero$&\bork
\ $d_2=\dtwo$\ \\\hline
\ $d_3=\dthree$\ &\bork
\ $d_5=\dfive$\ \\\hline
\ $d_{12}=-\frac1{48}(\frac{m-1}{m-2}\beta(m)-1)$&\bork
\ $d_{13}=\frac1{48}(1-\frac{4m-10}{m-2}\beta(m))$\\\hline
\ $d_{16}=\frac{17+5m}{192(m+1)}+\frac{23-2m-4m^2}
       {48(m-2)(m+1)}\beta(m)$&\bork
\ $d_{17}=-\frac{17+7m^2}{384(m^2-1)}+\frac{4m^3-11m^2+5m-1}{48
        (m^2-1)(m-2)}\beta (m)$\\\hline
\ $d_{20}=\frac1{8(m-3)}(\frac{5m-7}8-\frac{5m-9}3\beta(m))$&\bork
\ $d_{21}=\frac{m-1}{16(m-3)}(-1+2\beta(m))$\\\hline
\ $e_0 = \frac 1 {8(m-2)} \beta (m) $ & \bork
\ $e_1 = \frac 1 {8(m-1)(m-2)}\beta (m) $ \\\hline
\end{tabular}}

\medbreak\noindent{\bf Proof of (1).} We shall always choose a real localizing (or smearing)
function $F$. If the bundles $E_i$ and the data $(\gamma,\psi)$ are real, then $a_3$ is real.
Thus the coefficients $d_i$ are all real. Furthermore, since $D_{{\mathcal B}}$ is a
self-adjoint operator, the invariant $a_3$ is real in the general case.  Thus anti-Hermitian
invariants must appear with zero coefficient. By equation (\ref{ClifrelT}), $\gamma_a^T$ is
skew-Hermitian. We assumed $\Theta$ is Hermitian. Assertion (1) now follows as the following
terms are skew-Hermitian:
\begin{eqnarray*} 
   d_1 F[\ph \ph - \phc \phc ],\quad
   d_4 F[\gaa \ph \gaa \ph -\gaa \phc \gaa \phc],\quad
   d_7 F[\ph _{;m} -\phc _{;m}],\quad
   d_8 F[\gaa \ph _{:a} +\gaa \phc _{:a} ],\quad\\
   d_{11}FL_{aa} [ \ph - \phc],\quad
   d_{19}F_{;m} [\ph -\phc ],\quad
   e_2 F\gaa \Theta_{:a},\quad e_5 F\Theta [\ph -\phc ],\quad
   e_7 F\gaa \Theta \gaa [\ph -\phc ].
\end{eqnarray*}

\medbreak\noindent{\bf Proof of (2).} We consider the variation $\Theta (\varepsilon) := 
\Theta +\varepsilon$. For generic values of $\varepsilon$ the kernel of the associated
operator $A(\varepsilon)$ is trivial and the boundary condition remains unchanged and
thus the invariants $a_3(\varepsilon)$ are unchanged at these values of $\varepsilon$. The invariants
$a_3(\varepsilon)$ are locally computable. Thus $a_3$ is independent of $\varepsilon$. Assertion (2)
now follows from the identity:
\begin{eqnarray*}
0=\partial_\varepsilon a_3|_{\varepsilon=0}
  &=&\textstyle\int_{\partial M}\Trace\{2F(e_0+e_1\gamma_a^T\gamma_a^T)\Theta+Fe_3L_{aa}+
    F(e_4+e_6\gamma_a^T\gamma_a^T)(\hat\psi+\hat\psi^*)
   +e_8F_{;m}\}\\
  &=&\textstyle\int_{\partial M}\Trace\{2F(e_0-(m-1)e_1)\Theta+Fe_3L_{aa}+e_8F_{;m}
     +F(e_4-(m-1)e_6)(\hat\psi+\hat\psi^*)\}.
\end{eqnarray*}

\medbreak\noindent{\bf Proof of (3).}
We shall show that $\Trace\{W_{ab}\gaa\gab)=0$ and $\Trace\{W_{am}\gaa\}=0$ so these invariants
play no role. Note that $W_{ab}=-W_{ba}$. Furthermore, $[W,\gamma]=0$ as noted above. We use equation
(\ref{ClifrelT}) to compute:
\begin{eqnarray*}
&&\Trace\{W_{ab}\gaa\gab\}=\Trace\{\gaa W_{ab}\gab\}=\Trace\{W_{ab}\gab\gaa\}\text{ so}\\
&&\Trace\{W_{ab}\gaa\gab\}=\textstyle{1\over2}\Trace\{W_{ab}(\gaa\gab+\gab\gaa)\}
  =-\Trace\{W_{ab}\delta_{ab}\}=0.\end{eqnarray*}
Since $m\ne2$, we may show $\Trace\{W_{am}\gaa\}=0$ by computing
\begin{eqnarray*}
&&-(m-1)\Trace\{W_{am}\gaa\}=\Trace\{\gab\gab W_{am}\gaa\}=\Trace\{W_{am}\gab\gaa\gab\}\\
  &=&\Trace\{W_{am}(-2\delta_{ab}\gab-\gaa\gab\gab)\}=(-2+m-1)\Trace\{W_{am}\gaa\}.
\end{eqnarray*}

\medbreak\noindent{\bf Proof of (4).} We apply the local index theorem.
Let $M$ be the unit ball in ${\mathbb R}^m$ and let $E=E_1=E_2=Clif(M)$ be a trivial complex vector bundle of
dimension $2^m$ over $M$. Let $(\gamma,\nabla)$ be the standard Clifford module structure and flat connection on $E$.
Let $\psi_1$ be an arbitrary endomorphism of $E$ and set
$P_1:=\gamma^i\nabla_i+\psi_1:C^\infty(E_1)\rightarrow C^\infty(E_2)$; the formal adjoint is then given by
$P_2:=\gamma^i\nabla_i+\psi_1^*$ so
$\psi_2=\psi_1^*$. Let $D_1:=P_2P_1$ and $D_2:=P_1P_2$ with the appropriate boundary conditions ${\mathcal B}_i$. 
It follows from general principles that 
\begin{equation}
\Trace\{e^{-t(D_1)_{{\mathcal B}_1}}\}-\Trace\{e^{-t(D_2)_{{\mathcal B}_2}}\}
  =\text{index}(P_1,{\mathcal B}_1)\text{ so }
  a_3(D_1,{\mathcal B}_1)-a_3(D_2,{\mathcal B}_2)=0.\label{index}\end{equation} 
We use Lemma \ref{d9} (2) to identify the adjoint boundary conditions and $\Theta_2$
We use the equations of structure derived above and study the terms which are linear in
$\psi_1$ in equation (\ref{index}). Since $F=1$, Lemma \ref{d9} (1) shows the terms involving $d_9$ play no role. Thus:
\begin{eqnarray*}
&&\textstyle\int_{\partial M}\Trace\{d_6(-\gamma_m\psi_{1;m}+\psi_{1;m}^*\gamma_m)
     +(d_{10}L_{aa}+e_4\Theta_1+e_6\gamma_m\gamma_a\Theta_1\gamma_m\gamma_a)(-\gamma_m\psi_1+\psi_1^*\gamma_m)\}\\
&=&\textstyle\int_{\partial M}\Trace\{d_6(-\gamma_m\psi_{1;m}^*+\psi_{1;m}\gamma_m)
    +(d_{10}L_{aa}+e_4\gamma_m\Theta_1\gamma_m
      +e_6\gamma_m\gamma_a\gamma_m\Theta_1\gamma_m\gamma_m\gamma_a)(-\gamma_m\psi_1^*+\psi_1\gamma_m)\}\\
     &&\quad+\Trace\{e_4+(1-m)e_6)L_{aa}(-\gamma_m\psi_1^*+\psi_1\gamma_m)\}.
\end{eqnarray*}
The terms which are bilinear in $(\Theta_1,\psi_1)$ and $(\Theta_1,\psi_1^*)$ agree.
Since $e_4=(m-1)e_6$, the final term vanishes.
We set $\psi_1=f(x_m)\gamma_m$ to conclude that $d_6=0$ and that $d_{10}=0$. \qedbox

\medbreak{\noindent\bf Proof of (5).} We use the method of conformal variations described in
\cite{DGK}. Let $\tilde P$ be the Dirac operator on the upper hemisphere. Then $\tilde A$ is the
Dirac operator $S^{m-1}$. Since $S^{m-1}$ has a metric of positive scalar curvature,
$\ker(\tilde A)=\{0\}$ by the Lichnerowicz formula \cite{Lich}. We now perturb $\tilde P$ slightly to define
an operator of Dirac type $P_0$ on the ball which is formally self-adjoint. Let $A:=\frac12(B_0+B_0^*+L_{aa})$.
Since $A$ is close to $\tilde A$, $\ker A=\{0\}$ so the realization of $P$ is self-adjoint by Lemma \ref{d9}. Let
$f$ be a smooth function on
$M$. Let
\begin{eqnarray*}
&&ds^2(\varepsilon):=e^{2\varepsilon f}ds^2,\qquad\qquad\qquad\quad
dvol(\varepsilon)=e^{m\varepsilon f}dvol,\\
&&P(\varepsilon):=e^{-\frac{1+m}{2}\varepsilon f}P_0e^{-\frac{1-m}{2}\varepsilon f},\quad
P^*(\varepsilon):=e^{(-\frac{1-m}{2}-m)\varepsilon f}P_0e^{(m-\frac{1+m}{2})\varepsilon f}.
\end{eqnarray*}
We fix the metric on the bundle $E$. The metric determined by the leading symbol
of $P(\varepsilon)$ is $ds^2(\varepsilon)$ and $P(\varepsilon)$ is formally self-adjoint. We assume
$f=f(x_m)$ and $f|_{\partial M}=0$. Since $A(\varepsilon)-A_0=\frac{m-1}{2}\varepsilon
f_{;m}$ we set:
$$\Theta(\varepsilon)=\textstyle\frac{1-m}{2}\varepsilon f_{;m}+\textstyle\frac12L_{aa}(0)$$ 
to ensure that the boundary
conditions are unchanged. We use Lemma \ref{d9} and compute:
\begin{eqnarray*}
   L_{aa}(\varepsilon)&=&-\textstyle\frac12\partial_mg_{aa}(\varepsilon)=L_{aa}(0)+(1-m)\varepsilon f_{;m},\text{ and }\\
   \Theta_2(\varepsilon)&=&-\textstyle\frac{1-m}{2}\varepsilon f_{;m}-\textstyle\frac12L_{aa}(0)+L_{aa}(\varepsilon)
     =\Theta_1(\varepsilon)+\varepsilon(-2\frac{1-m}{2}+(1-m))f_{;m}\\
   &=&\Theta_1(\varepsilon).
\end{eqnarray*}
Let $\delta:=\partial_\varepsilon|_{\varepsilon=0}$. We compute
\begin{eqnarray}
  &&\delta\operatorname{Tr}_{L^2}\{e^{-tD(\varepsilon)}\}=
     -t\operatorname{Tr}_{L^2}\{\delta(
      D(\varepsilon))e^{-tD_0}\}
     =-2t\operatorname{Tr}_{L^2}\{\delta(P(\varepsilon))
        P_0e^{-tD_0}\}\nonumber\\
     &&\qquad=2t\operatorname{Tr}_{L^2}\{fD_0e^{-tD_0}\}
   =-2t\partial_t\operatorname{Tr}_{L^2}\{fe^{-tD_0}\}\text{. Consequently}\nonumber\\
   &&\delta a_3(1,D(\varepsilon),{\mathcal B})=
  (m-3)a_3(f,D_0,{\mathcal B}).\label{borko}\end{eqnarray}
We showed in \cite{DGK} that there exists a compatible family of unitary connections
${}^\varepsilon\nabla$ so that
$$\psi(\varepsilon)=\textstyle e^{-\varepsilon f}(\psi_0-\frac{m-1}{2}f_{;i}\gamma_i).$$
Since $\hat\psi_0(\varepsilon)=-\gamma_m\psi_0+\frac12(1-m)f_{;m}$, we have:
\begin{eqnarray*}
 \delta\hat\psi_0&=&\textstyle\frac{1-m}2f_{;m}=\delta\hat\psi_0^*,\\
 \delta d_0\Trace\{\hat\psi_0\hat\psi_0+\hat\psi_0\hat\psi_0\}
    &=&\textstyle\frac{1-m}{2}f_{;m}2d_0\Trace\{\hat\psi_0+\hat\psi_0^*\},\\
 \delta d_2\Trace\{\hat\psi_0\hat\psi_0^*\}&=&\textstyle\frac{1-m}{2}f_{;m}d_2\Trace\{\hat\psi_0+\hat\psi_0^*\},\\
 \delta d_3(\gamma_a^T\hat\psi_0\gamma_a^T\hat\psi_0+\gamma_a^T\hat\psi_0^*\gamma_a^T\hat\psi_0^*)
&=&\textstyle\frac{1-m}{2}f_{;m}2(1-m)d_3\Trace\{\hat\psi_0+\hat\psi_0^*\},\\
  \delta d_5(\gamma_a^T\hat\psi_0\gamma_a^T\hat\psi_0^*)&=&
\textstyle\frac{1-m}{2}f_{;m}(1-m)d_5(\hat\psi_0+\hat\psi_0^*).
\end{eqnarray*}
We use Lemma \ref{d9} to see $\delta\int_{\partial M}\Trace\{\gamma_a^T(\hat\psi_{0:a}-\hat\psi^*_{0:a})\}=0$. We use
computations from
\cite{BGa} to see
\begin{eqnarray*}
  \delta d_{12}\tau&=&d_{12}(-2(m-1)f_{;mm}+2(m-1)L_{aa}f_{;m}),\\
  \delta d_{13}\rho_{mm}&=&d_{13}(L_{aa}f_{;m}+(1-m)f_{;mm}),\\
  \delta d_{16}L_{ab}L_{ab}&=&-2d_{16}f_{;m}L_{aa}m\text{ and}\\
  \delta d_{17}L_{aa}L_{bb}&=&-2(m-1)d_{17}f_{;m}L_{aa}.
\end{eqnarray*}
We use equation (\ref{borko}) to see $\delta a_3(1,D(\varepsilon),{\mathcal B})+(3-m)a_3(f,D_0,{\mathcal B})=0$. 
We have $\delta\Theta=\frac12(1-m)f_{;m}$. Thus
\begin{eqnarray*}
\delta e_0\Trace\{\Theta^2\}&=&(1-m)e_0f_{;m}\Trace\{\Theta\},\\
\delta e_1\Trace\{\gamma_a^T\Theta\gamma_a^T\Theta\}&=&(1-m)(1-m)e_1f_{;m}\Trace\{\Theta\},\\
\delta e_4\Trace\{\Theta(\hat\psi_0+\hat\psi^*_0)\}&=&\textstyle\frac12(1-m)e_4f_{;m}\Trace\{\hat\psi_0+\hat\psi_0^*\}
      +(1-m)e_4f_{;m}\Trace\{\Theta\},\text{ and}\\
\delta e_6f_{;m}\Trace\{\gamma_a^T\Theta\gamma_a^T(\hat\psi_0+\hat\psi^*_0)\}
      &=&\textstyle\frac12(1-m)(1-m)e_6f_{;m}\Trace\{\hat\psi_0+\hat\psi_0^*\}
     +(1-m)(1-m)e_6f_{;m}\Trace\{\Theta\}.
\end{eqnarray*}
Since $e_0+(1-m)e_1=e_4+(1-m)e_6=0$, these terms play no role. Furthermore we have assumed $P=\gamma_i\nabla_i+\psi_0$
is self-adjoint. Thus $\psi_0=\psi_0^*$ and $\hat\psi_0+\hat\psi_0^*=-\gamma_m\psi_0+\psi_0\gamma_m$ and
$\Trace\{\hat\psi_0+\hat\psi_0^*)=0$. Thus this term yields no information. We complete the proof of assertion (5) by
computing:
\begin{eqnarray*}
0&=&\textstyle\int_{\partial M}(3-m)d_{18}f_{;m}\Trace\{\psi_0\}\\
  &&\quad+\{2(m-1)d_{12}+d_{13}-2d_{16}-2(m-1)d_{17}+(3-m)d_{20}\}f_{;m}\Trace\{L_{aa}\}\\
  &&\quad+\{-2(m-1)d_{12}+(1-m)d_{13}+(3-m)d_{21}\}\Trace\{f_{;mm}\}.
\end{eqnarray*}

\medbreak\noindent{\bf Proof of (6).} We now exploit the fact that the connection $\nabla$ is not canonically defined.
We let $M$ be the ball and let $E=E_1=E_2=Clif({\mathbb R}^m)\otimes V$ 
where $V$ is an auxiliary trivial vector
bundle. Let $\sigma_i:=I\otimes\tilde\sigma_i$ be skew-adjoint endomorphisms of $E$ commuting with the Clifford module
structure $\gamma$. Let 
$$\nabla_i(\varepsilon):=\nabla_i+\varepsilon\sigma_i$$
be a smooth $1$ parameter family
of unitary connections on $E$. Since $[\sigma_i,\gamma_j]=0$ for all $i,j$, we have $\nabla_i(\varepsilon)\gamma=0$ so
this is an admissible family of connections. We define
$$\psi(\varepsilon):=\psi_0-\varepsilon\gamma_j\sigma_j$$
to ensure that $P(\varepsilon)=\gamma_i\nabla_i(\varepsilon)+\psi(\varepsilon)=P$ is unchanged during the perturbation.
We have
\begin{eqnarray*}
   &&B(\varepsilon)=-\gamma_m(\gamma_a\nabla_a+\psi_0+\varepsilon\gamma_a\sigma_a-\varepsilon\gamma_i\sigma_i)
     =B_0-\varepsilon\sigma_m\text{ so}\\
   &&A(\varepsilon)
 =\textstyle\frac12(B(\varepsilon)+
  B(\varepsilon)^*)+\Theta (\epsilon)
=\textstyle\frac12(B_0+B_0^*)+\Theta_0=A_0.\\
\end{eqnarray*}
Thus the boundary conditions are unchanged by the perturbation if we set $\Theta(\varepsilon):=\Theta_0$.
Consequently, $a_3(F,D,{\mathcal B})$ is independent of the parameter $\varepsilon$. We compute
\begin{eqnarray*}
&&\delta\hat\psi(\varepsilon)=-\gamma_b^T\sigma_b-\sigma_m,\\
&&\delta\hat\psi(\varepsilon)^*=-\gamma_b^T\sigma_b+\sigma_m,\\
&&\delta d_0\Trace\{\hat\psi_0\hat\psi_0+\hat\psi_0^*\hat\psi_0^*\}
   =2d_0\Trace\{-\gamma_b^T\sigma_b(\hat\psi_0+\hat\psi_0^*)-\sigma_m(\hat\psi_0-\hat\psi_0^*)\},\\
&&\delta d_2\Trace\{\hat\psi_0\hat\psi_0^*\}
   =d_2\Trace\{-\gamma_b^T\sigma_b(\hat\psi_0+\hat\psi_0^*)+\sigma_m(\hat\psi_0-\hat\psi_0^*)\},\\
&&\delta d_3\Trace\{\gamma_a^T\hat\psi_0\gamma_a^T\hat\psi_0+\gamma_a^T\hat\psi_0^*\gamma_a^T\hat\psi_0^*\}
   =2d_3\Trace\{-\gamma_a^T\gamma_b^T\sigma_b\gamma_a^T(\hat\psi_0+\hat\psi_0^*)
     -\gamma_a^T\sigma_m\gamma_a^T(\hat\psi_0-\hat\psi_0^*)\}\\
   &&\qquad=2d_3\Trace\{(m-3)(-\gamma_b^T\sigma_b)(\hat\psi_0+\hat\psi_0^*)+(m-1)\sigma_m(\hat\psi_0-\hat\psi_0^*)\},\\
&&\delta d_5\Trace\{\gamma_a^T\hat\psi_0^*\gamma_a^T\hat\psi_0\}
     =d_5\Trace\{-\gamma_a^T\gamma_b^T\sigma_b\gamma_a^T(\hat\psi_0+\hat\psi_0^*)
       +\gamma_a^T\sigma_m\gamma_a^T(\hat\psi_0-\hat\psi_0^*)\}\\
   &&\qquad=d_5\Trace\{-(m-3)\gamma_b^T\sigma_b(\hat\psi_0+\hat\psi_0^*)+(1-m)\sigma_m(\hat\psi_0-\hat\psi_0^*)\},\\
&&\delta d_9F_{:a}\Trace\{\gamma_a^T(\hat\psi_0-\hat\psi_0^*)\}=d_9F_{:a}\Trace\{-2\gamma_a^T\sigma_m\}=0,\\
&&\delta e_4F\Trace\{\Theta(\hat\psi_0+\hat\psi_0^*)\}=-2e_4\Trace\{\Theta\gamma_b^T\sigma_b\},\text{ and}\\
&&\delta e_6F\Trace\{\Theta\gamma_a^T(\hat\psi_0+\hat\psi_0^*)\gamma_a^T\}
   =-2e_6\Trace\{\Theta(m-3)\gamma_b^T\sigma_b\}.\\
\end{eqnarray*}
This yields the relation:
\begin{eqnarray*}
0=\textstyle\int_{\partial M}&\{&2d_0+d_2+(m-3)(2d_3+d_5)\}\Trace\{-\gamma_b^T\sigma_b(\hat\psi_0+\hat\psi_0^*)\}\\
+&\{&-2d_0+d_2+(m-1)(2d_3-d_5)\}\Trace\{\sigma_m(\hat\psi_0-\hat\psi_0^*)\}\\
+&\{&-2e_4-2(m-3)e_6\}\Trace\{\Theta\gamma_b^T\sigma_b\}.\\
\end{eqnarray*}

To determine $d_9$ we extend the setting to an endomorphism valued
smearing function. We study those terms which involve the tangential
covariant derivatives of $F$. After taking into account the lack of
commutativity, we see that these terms take the form:
$$\{u_1\trace(F_{:a}\gamma_a^T(\hat\psi-\hat\psi^*)),
    u_2\trace(F_{:a}\gamma_a^T(\hat\psi+\hat\psi^*)),
    u_3\trace(F_{:a}(\hat\psi-\hat\psi^*)\gamma_a^T),
    u_4\trace(F_{:a}(\hat\psi+\hat\psi^*)\gamma_a^T),
    u_5\trace(F_{:a}\gamma_a^T\theta),
    u_6\trace(F_{:a}\theta\gamma_a^T\}.$$
If $F$ is then taken to be scalar, we see that $d_8=-u_2-u_4$,
$d_9=-u_1-u_3$, and $e_2=-u_5-u_6$.  We set $\psi_0=0$, $\theta=0$, and
$\sigma_a=0$. Then $\delta (\hat\psi-\hat\psi^*)=-2\sigma_m$  and
$\delta(\hat\psi+\hat\psi^*)=0$. Since $\sigma_m$ commutes with
$\gamma_a^T$, we get
$$0=-2(u_1+u_3)\trace(F_{:a}\gamma_a^T\sigma_m)$$
since these are the only terms in the variation involving the covariant
derivatives of $F$. (As $\trace(\gamma_a^T\sigma_m)=0$, it is necessary
to take $F_{:a}$ endomorphism valued for this argument to work). We can
now conclude that $u_1+u_3=0$. This shows $d_9=0$ and completes the
proof of assertion 6d). \qedbox

\medbreak\noindent{\bf Proof of (7).} As in the proof of (5), let $P_0$ be a small perturbation of the
Dirac operator on the upper hemisphere so that
$\ker(A_0)=\{0\}$ where $A_0:=\frac12(B_0+B_0^*+L_{aa})$; the realization of
$P$ is self-adjoint. We consider a variation of the form $P(\varepsilon):=P+\varepsilon$. We then have
$B(\varepsilon)=B_0-\gamma_m\varepsilon$ and thus
$A(\varepsilon)=\frac12(B(\varepsilon)+B^*(\varepsilon)+L_{aa})
=A_0$ is independent of the parameter $\varepsilon$. Thus $P(\varepsilon)$ is
self-adjoint. If
$\{\phi_k,\lambda_k\}$ is a spectral resolution of
$P$, then $\{\phi_k,\lambda_k+\varepsilon\}$ will be a spectral resolution of $P(\varepsilon)$. We compute:
\begin{eqnarray*}
     &&\textstyle\sum_k\partial_\varepsilon^2\{a_k(1,P(\varepsilon)^2,{\mathcal B})\}|_{\varepsilon=0}t^{(n-m)/2}
    \sim\partial_\varepsilon^2\Trace\{e^{-t(P+\varepsilon)^2}\}|_{\varepsilon=0}\\
   &=&\partial_\varepsilon\Trace\{-2t(P+\varepsilon)e^{-tP(\varepsilon)^2}\}|_{\varepsilon=0}
    =\Trace\{(-2t+4t^2P^2)e^{-tP^2}\}\\
  &=&-2t\Trace\{(1+2t\partial_t)e^{-tP^2}\}
  \sim-2t\textstyle\sum_n\{1+(n-m)\}a_n(1,P^2,{\mathcal B})t^{(n-m)/2}.
\end{eqnarray*}
We take $(k,n)=(3,1)$ and equate the coefficient of $t^{(3-m)/2}$ in the two expansions to
see:
\begin{equation}\partial_\varepsilon^2a_3(1,P(\varepsilon)^2,{\mathcal B})
   =-2(2-m)a_1(1,P^2,{\mathcal B})\label{borkXX}
\end{equation}
We use Theorem \ref{THM1} to see
\begin{equation}
 a_1(1,P^2,{\mathcal B})
=(4\pi)^{-(m-1)/2}\textstyle\frac14(\beta(m)-1)\textstyle\int_{\partial M}\Trace\{I\}\label{borkXY}
\end{equation}
We have $\hat\psi(\varepsilon)=\hat\psi_0-\gamma_m\varepsilon$ and $\hat\psi(\varepsilon)^*=\hat\psi_0+\gamma_m\varepsilon$.
Assertion (7) now follows from equations (\ref{borkXX}), (\ref{borkXY}), and the following identity:
$$
\partial_\varepsilon^2a_3(1,P(\varepsilon)^2,{\mathcal B})
   =(4\pi)^{-(m-1)/2}\textstyle\int_{\partial M}\{-4d_0+2d_2-4(m-1)d_3+2(m-1)d_5\}\Trace\{I\}.\ \qedbox
$$

\medbreak\noindent{\bf Proof of (8).} As in the proof of (5), let $P_0$ be a small perturbation of the
Dirac operator on the upper hemisphere so that $P_0$ is formally self-adjoint and so that
$\ker(A_0)=\{0\}$ where 
$A_0:=\frac12(B_0+B_0^*+\Theta_0+L_{aa})$. We assume that the realization of
$P_0$ is self-adjoint. We consider a variation of the form
\def\vep{\sqrt{-1}\varepsilon} $P(\varepsilon):=P_0+\vep$. Then
$$\hat\psi(\varepsilon)=\hat\psi_0-\vep\gamma_m, \quad
\hat\psi^*(\varepsilon)=\hat\psi_0^*-\vep\gamma_m,\text{ so we set }
 \Theta(\varepsilon)=\Theta_0+\vep\gamma_m.$$
Then $A(\varepsilon)=A_0$ so the boundary condition is
unchanged. Thus
$P^*(\varepsilon ) =P_0-\vep$ and
$D=P^2+\varepsilon^2$. Consequently we have
\begin{equation}
\Trace\{e^{-tD(\varepsilon)}\}=e^{-t\varepsilon^2}\Trace\{e^{-tD_0}\}\text { so }
  a_3(1,D(\varepsilon),{\mathcal B})=a_3(1,D_0,{\mathcal B})-\varepsilon^2a_1(1,D_0,{\mathcal B}).\label{borkzz}
\end{equation}
We compute:
\begin{eqnarray*}
d_0\Trace\{\hat\psi\hat\psi+\hat\psi^*\hat\psi^*\}(\varepsilon)
  &=&d_0\Trace\{\hat\psi_0\hat\psi_0+\hat\psi_0^*\hat\psi_0^*\}-2d_0\vep\Trace\{\gamma_m(\hat\psi_0+\hat\psi_0^*)\}
     +2d_0\varepsilon^2\Trace\{I\}\\
d_2\Trace\{\hat\psi\hat\psi^*\}(\varepsilon)
  &=&d_2\Trace\{\hat\psi_0\hat\psi_0^*\}-d_2\vep\Trace\{\gamma_m(\hat\psi_0+\hat\psi_0^*)\}+d_2\varepsilon^2\Trace\{I\}\\
d_3\Trace\{\gamma_a^T\hat\psi\gamma_a^T\hat\psi
+\gamma_a^T\hat\psi^*\gamma_a^T\hat\psi^*\}(\varepsilon)
  &=&d_3\Trace\{\gamma_a^T\hat\psi_0\gamma_a^T\hat\psi_0
+\gamma_a^T\hat\psi_0^*\gamma_a^T\hat\psi_0^*\}
   -2d_3(m-1)\vep\Trace\{\gamma_m(\hat\psi_0+\hat\psi_0^*)\}
   \\&&\qquad\qquad+2(m-1)d_3\varepsilon^2\Trace\{I\}\\
d_5\Trace\{\gamma_a^T\hat\psi\gamma_a^T\hat\psi^*\}(\varepsilon)
  &=&d_5\Trace\{\gamma_a^T\hat\psi_0\gamma_a^T\hat\psi_0^*\}
   -d_5(m-1)\vep\Trace\{\gamma_m(\hat\psi_0+\hat\psi_0^*)\}
   \\&&\qquad\qquad+(m-1)d_5\varepsilon^2\Trace\{I\}\\
e_0\Trace\{\Theta\Theta\}(\varepsilon)
  &=&e_0\Trace\{\Theta_0\Theta_0\}+2e_0\vep\Trace\{\gamma_m\Theta_0\}+e_0\varepsilon^2\Trace\{I\}\\
e_1\Trace\{\gamma_a^T\Theta\gamma_a^T\Theta\}(\varepsilon)
  &=&e_1\Trace\{\gamma_a^T\Theta_0\gamma_a^T\Theta_0\}+2e_1(m-1)\vep\Trace\{\gamma_m\Theta_0\}
    +e_1(m-1)\varepsilon^2\Trace\{I\}\\
e_4\Trace\{\Theta(\hat\psi+\hat\psi^*)\}(\varepsilon)
  &=&e_4\Trace\{\Theta_0(\hat\psi_0+\hat\psi_0^*)\}+e_4\vep\Trace\{\gamma_m(\hat\psi_0+\hat\psi_0^*-2\Theta_0)\}
    -2e_4\varepsilon^2\Trace\{I\}\\
e_6\Trace\{\gamma_a^T\Theta\gamma_a^T(\hat\psi+\hat\psi^*)\}(\varepsilon)
 &=&e_6\Trace\{\gamma_a^T\Theta_0\gamma_a^T(\hat\psi_0+\hat\psi_0^*)
      +e_6(m-1)\vep\Trace\{\gamma_m(\hat\psi_0+\hat\psi_0^*-2\Theta_0)\}
    \\&&\qquad\qquad-2e_6(m-1)\varepsilon^2\Trace\{I\}.
\end{eqnarray*}
Thus we have
\begin{eqnarray*}
0&=&\textstyle\int_{\partial
M}\{-2d_0-d_2-2d_3(m-1)-d_5(m-1)+e_4+(m-1)e_6\}\sqrt{-1}\Trace\{\gamma_m(\hat\psi_0+\hat\psi_0^*)\}\\
  &\{&2e_0+(m-1)e_1-2e_4-2(m-1)e_6\}\sqrt{-1}\Trace\{\gamma_m\Theta_0\}.
\end{eqnarray*}
To ensure that $P_0$ is self-adjoint, we must have $\gamma_m\Theta_0\gamma_m=\Theta_0+L_{aa}$. Thus, in particular
$\Trace\{\gamma_m\Theta_0\}=0$. Furthermore $\psi_0=\psi_0^*$. Thus
$\Trace\{\gamma_m(\hat\psi_0+\hat\psi_0^*)\}=\Trace\{\gamma_m(-\gamma_m\psi_0+\psi_0\gamma_m)\}=0$. 
Consequently the coefficient of $\varepsilon$ produces no information. We use equation (\ref{borkzz}) to identify the
coefficient of $\varepsilon^2$ and see
\begin{eqnarray*}
&&(4\pi)^{-(m-1)/2}\textstyle\int_{\partial M}\Trace\{I\}\cdot\{2d_0+d_2+2(m-1)d_3+(m-1)d_5\\
 &&\qquad\qquad\qquad\qquad+e_0+e_1(m-1)-2e_4-2e_6(m-1)\}
  \\
&=&-(4\pi)^{-(m-1)/2}\textstyle\int_{\partial M}\textstyle\frac14(\beta(m)-1)\Trace\{I\}.\ \qedbox
\end{eqnarray*}
\medbreak\noindent{\bf Proof of (9).} Grubb and Seeley \cite{GSc} gave a complete description of the singularities of
$\Gamma(s)\Trace\{FD_1^{-s}\}$ in the cylindrical case - i.e. when the structures are product near the boundary (see Theorem
2.1 \cite{GSc} for details). We use the inward geodesic flow to identify a neighborhood of the boundary
$\partial M$ in $M$ with the collar ${\mathcal C}=\partial M\times(-\epsilon,0]$. Let $(y,x_m)$ be coordinates on
${\mathcal C}$. We suppose that $P=\gamma_m(\partial_m+A)$ on ${\mathcal C}$ where $A$ is a tangential self-adjoint operator
of Dirac type whose coefficients are independent of the normal variable $x_m$. Thus $A=\gamma_a^T\nabla_a+\hat\psi$ where
$\hat\psi$ is self-adjoint. Since
$d_9$ vanishes we may take $F=1$.
We use Equation (13) \cite{DGK} to see that:
\begin{equation}a_3(F,D,{\mathcal B})=\frac14
\left(\frac{m-1}{m-2}\beta(m)-1\right)a_2(F,A^2).\label{equationB}
\end{equation}
We use Theorem 4.1 \cite{BGb} to see that:
\begin{equation}a_2(F,A^2)=-\textstyle\frac1{12}(4\pi)^{-(m-1)/2}\textstyle\int_{\partial M}F
\Trace\{R_{abba}+(12-6(m-1))\hat\psi\hat\psi+6\gamma_a^T\hat\psi\gamma_a^T\hat\psi\}.\label{equationC}
\end{equation}
Assertion (9) now follows from equations (\ref{equationB}), (\ref{equationC}) and the computation:
\begin{eqnarray*}
a_3(F,D,{\mathcal B})=(4\pi)^{-(m-1)/2}\textstyle\int_{\partial M} F[
(2d_0+d_2)\Trace\{\hat\psi\hat\psi\}+(2d_3+d_5)
\Trace\{\gamma_a^T\hat\psi\gamma_a^T\hat\psi\}+d_{12}\Trace\{I\}].
\ \qedbox
\end{eqnarray*}

\medbreak\noindent{\bf Proof of (10).} This follows from 
computations on the ball. We follow the description in \cite{DGK} and
extend the results to the ones needed for $a_3$. 
If $r\in[0,1]$ is the radial normal coordinate and if $d\Sigma^2$ is
the usual metric on the unit sphere $S^{m-1}$, then
$ds^2=dr^2+r^2d\Sigma^2$.
The inward unit normal on the boundary is $-\partial_r$. The only
nonvanishing components of the Christoffel symbols are
\begin{eqnarray}
\Gamma_{abc}= \frac 1 r \tilde{\Gamma}_{abc}\text{ and }
\Gamma_{abm}= \frac 1 r \delta_{ab};\nn
\end{eqnarray}
the second fundamental form is given by $L_{ab}=\delta_{ab}$.
We denote by
$\tilde{\Gamma}_{abc}$ the Christoffel symbols associated with the
metric $d\Sigma^2$ on the sphere 
$S^{m-1}$ and tilde will always refer to this metric.

We will consider the Dirac operator $P=\gamma^\nu\partial_\nu$ on the 
ball; we take the flat connection $\nabla$ and set
$\psi=0$. We suppose $m$ even (there is a corresponding
decomposition for $m$ odd) and use the following representation of the 
$\gamma$-matrices:
\begin{eqnarray}
&&\gamma_{a(m)}=\left(
   \begin{array}{cc}
               0 &  \sqrt{-1}\cdot \gamma_{a(m-1)}    \\
      -\sqrt{-1}\cdot \gamma_{a(m-1)}    &     0
    \end{array}    \right)\text{ and }\nn\\
\quad
&&\gamma_{m(m)} = \left(
     \begin{array}{cc}
         0       &    \sqrt{-1}\cdot 1_{m-1}   \\
    \sqrt{-1}\cdot 1_{m-1}\quad\    &      0
    \end{array}   \right)   .\nonumber
\end{eqnarray}
We stress that $\gamma_{j(m)}$ are the $\gamma$-matrices projected
along some vielbein system $e_j$. Decompose $\nabla_j = e_j + \omega_j$ where
$\omega_j=\frac 1 4 \Gamma_{jkl} \gamma_{k(m)} 
\gamma_{l(m)}$ is the connection $1$ form
of the spin connection. Note that
$$
\nabla_a = \frac 1 r\left( \left(
        \bea {cc}
        \tilde{\nabla}_a & 0 \\
         0 & \tilde{\nabla}_a
         \eea  \right)  +\frac 1 2 \gamma_{a(m)}^T\right).
$$
Let $\tilde P$ the Dirac operator on the sphere. We have:
\begin{eqnarray}
&&P=\left(\frac{\partial}{\partial x_m}-\frac{m-1}{2r} \right) \gamma_{m(m)}
         +\frac 1 r \left(
\begin{array}{cc}
       0   & \sqrt{-1} \tilde P \\
    -\sqrt{-1} \tilde P  & 0
\end{array}   \right).\nn
\end{eqnarray}
Let $d_s$ be the dimension of the spin bundle on the disk; $d_s=2^{m/2}$ if $m$ is even.
The spinor modes ${\mathcal Z} _\pm ^{(n)}$ on the sphere are discussed in
\cite{roberto}. We have
\begin{eqnarray*}
&&\tilde P {\mathcal Z} _\pm ^{(n)} (\Omega )=  \pm \left( n+\frac{m-1} 2
\right)
             {\mathcal Z} _\pm ^{(n)} (\Omega )\text{ for }n=0,1,...;\\
&&  d_n(m):=\dim {\mathcal Z} _\pm ^{(n)} (\Omega )= \frac 1 2 d_s \left(
    \bea {c}
       m+n-2 \\
        n
      \eea \right) .
\end{eqnarray*}
Let $J_{\nu} (z)$ be the Bessel functions. These satisfy the
differential equation \cite{grad}:
\begin{eqnarray*}
\frac{d^2 J_{\nu} (z)} {dz^2} +\frac 1 z \frac{dJ_{\nu} (z)} {dz} +
\left( 1-\frac{\nu^2}{z^2} \right) J_{\nu }(z) =0.
\end{eqnarray*}
Let  $P\varphi_\pm = \pm \mu \varphi_\pm$ be an eigen function
of $P$. Modulo a suitable radial normalizing constant $C$, we may express:
\begin{eqnarray}
\varphi_{\pm}^{(+)}&=&{\frac{C}{r^{(m-2)/2}}} \left(
     \begin{array}{c}
        iJ_{n+m/2}(\mu r)
       \,Z^{(n)}_+(\Omega ),  \\
     \pm J_{n+m/2-1}(\mu r)\,Z^{(n)}_+(\Omega )
        \end{array}  \right) , \text{ and} \label{eq2.59}\\
\varphi_{\pm}^{(-)}&=&{\frac{C}{r^{(m-2)/2}}}\left(
     \begin{array}{c}
     \pm J_{n+m/2-1}(\mu r)\,Z^{(n)}_
-(\Omega )  \\
   iJ_{n+m/2}(\mu r)\,Z^{(n)}_-(\Omega ) \end{array}
\right).
\label{solutions}
\end{eqnarray}
Let $\nabla_a^T:=\nabla_a- \frac 1 2 L_{ab}\gamma_{b(m)}^T$. Then 
$\nabla^T$ is a compatible unitary connection for the induced
Clifford modules structure $\gamma^T$; see \cite{G} for details.
The tangential operator $B$ takes the form:
\begin{eqnarray}
B &=&\gamma_{a(m)}^T \left(\nabla_a^T+\frac 1 2 L_{ab} \gamma_{b(m)}^T 
\right) = \left(
\begin{array}{cc}
       -\tilde P -\frac {m-1} 2 & 0 \\
   0 & \tilde P -\frac {m-1} 2
 \end{array} \right).\nn
\end{eqnarray}
We have in particular $B=B^*$. 
We take $\Theta = \frac{m-1} 2 \,\, 1_m$. The operator $A$ used to define 
spectral boundary conditions then reads 
\begin{eqnarray}
A=\left(
\begin{array}{cc}
-\tilde P & 0 \\
0 & \tilde P
\end{array} \right). \nn
\end{eqnarray}
The eigenstates and eigenvalues of $A$ then are easily determined:
\beq
A\left(\bea {c}
          {\mathcal Z}_+^{(n)}(\Omega ) \\
            {\mathcal Z}_-^{(n)} ( \Omega )
        \eea \right) &=& -\left( n+\frac{m-1} 2 \right)
\left(\bea {c}
          {\mathcal Z}_+^{(n)} (\Omega )\\
            {\mathcal Z}_-^{(n)} (\Omega )
        \eea \right)\text{ and} \nn\\
A\left(\bea {c}
          {\mathcal Z}_-^{(n)} (\Omega )\\
            {\mathcal Z}_+^{(n)} (\Omega )
        \eea \right) &=&  \left( n+\frac{m-1} 2 \right)
\left(\bea {c}
          {\mathcal Z}_-^{(n)} (\Omega )\\
            {\mathcal Z}_+^{(n)} (\Omega )
        \eea \right)\text{ for }n=0,1,....\nn
\eeq
The boundary condition suppresses the
non-negative spectrum of $A$. Applying the boundary conditions on the 
solutions 
(\ref{eq2.59}) and (\ref{solutions}), we see that
the non-negative modes of $A$ are associated with the radial factor
$J_{n+\frac m 2 -1} (\mu  r)$. Hence the implicit eigenvalue
equation is
\beq
J_p (\mu ) =0\text{ where }p=n+\frac m 2-1.\label{implicit}
\eeq
In \cite{bek,cmp,jon,anageo} 
a method has been developed for calculating the 
associated heat-kernel coefficients for smearing (or localizing) function $F=1$; 
in \cite{smear} this has been generalized to $F=F(r)$. 
We summarize the essential results from these papers briefly;
in principal one could calculate any number of coefficients.
We first suppose that
$F=1$. Instead of looking directly at the heat-kernel we will consider the
zeta-function $\zeta (s)$ of the operator $P^2$ and use
the relationship 
between the pole structure of the zeta function and the
asymptotics of the heat equation:
\beq
a_k =  \mbox{Res }_{s=\frac {m-k}2}\Gamma(s)\zeta (s).\label{ex1}
\eeq
Thus to compute $a_3$, we must determine the residues of the
zeta-function $\zeta (s)$ at the value $s=(m-3)/2$. We use the eigenvalue equation
(\ref{implicit}) to express
\beq
\zeta (s) = 4 \sum_{n=0}^\infty d_n (m)
\int_{\mathcal C} \frac{dk}{2\pi i} k^{-2s} \frac \partial {\partial k} \ln
J_p (k) ,
\label{ex2}
\eeq
where the contour ${\mathcal C}$ runs counterclockwise and 
encloses all the solutions
of (\ref{implicit}) which lie on the positive real axis. 
The factor of four
comes from the four types of solutions in (\ref{eq2.59}) and (\ref{solutions}).
The representation equation (\ref{ex2}) is well defined only for
$\Re s > m/2$, so the first task is to construct the analytical continuation to
the left. In order to do that, it is convenient to 
define a modified zeta function
\beq
\zeta ^{(n)} (s) =
\int_{\mathcal C} \frac{dk}{2\pi i} k^{-2s} \frac \partial {\partial k}
\ln k^{-p} J_p (k)  ;\nn
\eeq
the additional factor $k^{-p}$ has been introduced to avoid
contributions coming from the origin. Since no additional pole is enclosed,
the integral is unchanged.

It is the behaviour of $\zeta^{(n)} (s)$ 
as $n\to \infty$ which controls the convergence of the
sum over $n$. The different orders in $n$ can be
studied by shifting the contour to the imaginary axis and by using the
uniform asymptotic expansion of the resulting Bessel function $I_p (k)$.
To ensure that the resulting expression converges for some range of
$s$ when shifting the contour 
to the imaginary axis, we add a small positive constant
to the eigenvalues. For $s$ in the strip $1/2 < \Re s <1$, we have:
\beq
\zeta ^{(n)} (s) = \frac{\sin (\pi s)} \pi
\int_\epsilon^\infty dk (k^2-\epsilon^2)^{-s}  \frac \partial {\partial k}
\ln k^{-p} I_p (k).\nn
\eeq
We introduce some additional notation dealing with the uniform asymptotic expansion of
the Bessel function. For $p\to\infty$ with $z=k/p$ fixed, we use results of
\cite{abra} to see that:
\begin{eqnarray}
&&I_p (zp) \sim \frac 1 {\sqrt{2\pi p}} \frac{e^{p\eta}}{(1+z^2)^{1/4}}
\left[ 1+\sum_{l=1}^\infty \frac{u_l (t)} { p^l} \right]\text{ where}\label{ex4}\\
&&t=1/\sqrt{1+z^2}\text{ and }\eta = \sqrt{1+z^2}+\ln [z/(1+\sqrt{1+z^2})].\nn
\end{eqnarray}
Let $u_0(t)=1$. We use the recursion relationship given in \cite{abra} to determine
the polynomials $u_l (t)$ which appear in equation (\ref{ex4}):
\beq
u_{l+1} (t) = \frac 1 2 t^2 (1-t^2) u_l' (t) +\frac 1 8 \int_0^t d\tau
(1-5\tau^2) u_l (\tau).\nn
\eeq
We also need the coefficients $D_m (t)$ defined by the cumulant expansion:
\beq
\ln \left[  1+\sum_{l=1}^\infty \frac{u_l (t)}{p^l} \right]
\sim \sum_{q=1}^\infty \frac{D_q (t)}{p^q} .\label{ex5}
\eeq
The eigenvalue multiplicities $d_n (m)$ are ${\mathcal O} (n^{m-2})$
as $n\to \infty$. 
Consequently, the leading behaviour of every term is of the order of
$p^{-2s-q+m-2}$; thus on the half plane $\Re s > (m-4)/2$, only the values
$q=1$ and $q=2$ contribute to the residues of the zeta-function. We have
\beq
D_1 (t) &=&\frac 1 8 t -\frac 5 {24} t^3,\text{ and }
D_2 (t)= \frac 1 {16} t^2 -\frac 3 8 t^4 +\frac 5 {16} t^6 . \nn
\eeq
We use equation (\ref{ex4}) to decompose
\beq
\zeta^ {(n)} (s) &=& A_{-1} ^{(n)} (s) + A_0 ^{(n)} (s) +
A_1 ^{(n)} (s) + R^{(n)} (s),\text{ where} \nn\\
A_{-1}^{(n)} (s) &=&  \frac{\sin \pi s}{\pi} \int_{\epsilon / p} ^\infty
dz [(zp)^2 -\epsilon^2] ^{-s} \frac {\partial}{\partial z} \ln
\left( z^{-p} e^{p\eta} \right) ,\nn\\
A_{0 }^{(n)} (s) &=&  \frac{\sin \pi s}{\pi} \int_{\epsilon / p} ^\infty
dz [(zp)^2 -\epsilon^2] ^{-s} \frac {\partial}{\partial z} \ln
\left( 1+z^2 \right)^{-1/4} ,\nn\\
A_q ^{(n)} (s) &=&  \frac{\sin \pi s}{\pi} \int_{\epsilon / p} ^\infty
dz [(zp)^2 -\epsilon^2] ^{-s} \frac {\partial}{\partial z}
\left(\frac{D_q (t) } {p^q} \right).\nn
\eeq
The remainder $R^{(n)}(s)$ is such that $ \sum_{n=0}^\infty d_n (m) R^{(n)} (s)$
is analytic on the half plane $\Re s > (m-4)/2$.

Let ${_2F_1}$ be the hypergeometric function. We have
\begin{eqnarray*}
&&{_2F_1} (a,b;c;z) = \frac{\Gamma (c) }{\Gamma (b) \Gamma (c-b)}
\int_0^1 dt t^{b-1} (1-t)^{c-b-1} (1-tz)^{-a},\text{ and}\\
&& \int_{\epsilon /p } ^\infty dz \,\, [(zp)^2 -\epsilon ^2] ^{-s} \frac
{\partial}{\partial z} t^l=
-\frac l 2 \frac {\Gamma (s+\frac l 2) \Gamma (1-s)}{\Gamma (1+\frac l 2) } p^l
[\epsilon^2 + p^2] ^{-s-l/2} .\nn
\end{eqnarray*}
We use the first identity to study $A_{-1}^{(n)}(s)$ and $A_0^{(n)}(s)$; we
use the second identity to study $A_1^{(n)}(s)$ and $A_2^{(n)} (s)$. 
This shows that
\beq
A_{-1}^{(n)} (s) &=&  \frac{\epsilon^ {-2s +1} } {2\Gamma( \frac{1}{2}) }
\frac{\Gamma (s-\frac 1 2)} {\Gamma (s)} {_2F_1}
(-{\frac{1}{2}},s-{\frac{1}{2}};
{\frac{1}{2}};-({\frac p \epsilon}
)^2 ) \nn  -\frac p 2
\epsilon ^{-2s} \nn
\\ A_0 ^{(n)} (s) &=&  -\frac 1 4 (p^2 + \epsilon ^2) ^{-s} , \nn \\
A_1 ^{(n)} (s) &=&  \frac 1 8 \frac 1 {\Gamma (s)} \left[
             -\frac{\Gamma (s+\frac{1}{2})}{\Gamma( \frac{1}{2}) }
            (p^2 +\epsilon ^2)^{-s-\frac{1}{2}}  \right]
                \nn\\
   & & -\frac 5 {24} \frac 1 {\Gamma (s)} \left[
-2\frac{\Gamma (s+\frac 3 2)} {\Gamma( \frac{1}{2}) }
 p^2 (p^2 +\epsilon ^2 )^{-s-\frac 3 2} \right],\nn\\
A_2^{(n)} (s) &=& \frac 1 {16} \frac 1 {\Gamma (s)} \left[ 
  -\Gamma (s+1) (p^2+\epsilon^2)^{-s-1}\right] \nn\\
  & &-\frac 3 8 \frac 1 {\Gamma (s)} \left[
       -\Gamma (s+2) p^2 (p^2+\epsilon^2)^{-s-2}\right] \nn\\
  & &+\frac 5 {16} \frac 1 {\Gamma (s)} \left[ -\frac 1 2 
\Gamma (s+3) p^4 (p^2+\epsilon^2 )^{-s-3}\right] . \nn
\eeq
In the limit $\epsilon \to 0$, the resulting zeta-function
which appears is connected to the spectrum on the sphere. Let $d:=m-1$.
We define the
base zeta-function $\zeta_{S^d}$ and the Barnes zeta-function \cite{barnes} $\zeta_{{\mathcal B}}$,
\begin{eqnarray*}
&& \zeta _{S^d} (s) = 4 \sum_{n=0}^\infty d_n (m) p^{-2s}\text{ and }
\zeta_{{\mathcal B}} (s,a) = 
\sum_{n=0}^\infty d_n (m) (n+a)^{-s}.\end{eqnarray*}
We then have the relation
$\zeta_{S^d} (s) = 2d_s \zeta_{{\mathcal B}} \left( 2s, \frac m 2 -1 \right)$.
For $i=-1$, $i=0$, $i=1$ and $i=2$, 
we shall define $A_i (s) = 4\sum_{n=0}^\infty d_n (m) A_i ^{(n)} (s)$.
We take the limit as $\epsilon \to 0$ to see that
\beq
&&A_{-1} (s)=   \frac 1 {4\Gamma( \frac{1}{2})} \frac {\Gamma (s-\frac 1 2) }{\Gamma
(s+1)}
\zeta_{S^d} (s-\frac 1 2) , \label{ex11} \\
&&A_0 (s)=   -\frac 1 4 \zeta_{S^d} (s) , \label{ex12} \\
&&A_1 (s)= -\frac 1 {\Gamma (s) } \zeta_{S^d} (s+\frac 1 2) \left[
   \frac 1 {8\Gamma( \frac{1}{2})} \Gamma (s+\frac 1 2) -
  \frac 5 {12 \Gamma( \frac{1}{2}) } \Gamma (s+\frac3 2) \right], 
      \label{ex13}\\
&&A_2 (s) = -\frac 1 {\Gamma (s)} \zeta_{S^d} (s+1) \left[ 
     \frac 1 {16} \Gamma (s+1) -\frac 3 8 \Gamma (s+2) +\frac 5 {32} 
      \Gamma (s+3) \right] . \label{ex13a} 
\eeq
We used the Mellin-Barnes integral representation of the
hypergeometric functions \cite{grad} to
calculate $A_{-1} (s)$:
\beq
{_2F_1} (a,b;c;z) = \frac{\Gamma (c)}{\Gamma (a) \Gamma (b)} \frac 1 {2\pi i}
\int_{{\mathcal C}} dt \,\, \frac{\Gamma (a+t) \Gamma (b+t) \Gamma (-t) }
{\Gamma (c+t)} (-z)^t.\label{hyper}
\eeq
The contour of integration is such that 
the poles of $\Gamma (a+t) \Gamma (b+t) /
\Gamma (c+t)$ lie to the left of the contour and so that the poles of $\Gamma (-t)$ lie to the
right of the contour. We stress that before interchanging the sum and the integral, we must
shift the contour ${\mathcal C}$ over the pole at $t=1/2$ to the left; this cancels
the term  $-\frac p 2 \epsilon ^{-2s}$ appearing in the 
expression for $A_{-1}$ above.

This reduces the analysis of the
zeta function on the ball to analysis of a zeta function on the boundary. We compute the
residues of
$\zeta (s)$ from the residues of $\zeta _{{\mathcal B}}(s,a)$. To compute
these residues, we first express 
$\zeta_{{\mathcal B}} (s,a)$ as a contour integral. Let
${\mathcal C}$ be the Hankel contour.
\beq
\zeta_{{\mathcal B}} (s,a)
&=& \sum_{n=0}^\infty \left(
\begin{array}{c}
   d+n-1 \\
   n
\end{array}
\right) (n+a)^{-s} = \sum_{\vec m \in \nats_0^d} (a+m_1+...+m_d)^{-s}
\nn\\
&=& \frac{\Gamma (1-s) }{2\pi} \int_{{\mathcal C}} dt \,\, (-t)^{s-1}
\frac{e^{-at}} {(1-e^{-t})^d}.\nn
\eeq
The residues of $\zeta_{{\mathcal B}} (s,a)$
are intimately connected with the generalized Bernoulli polynomials
\cite{norlund},
\beq
\frac{e^{-at} } {(1-e^{-t} )^d} = (-1)^d
\sum_{n=0} ^\infty \frac{(-t)^{n-d} } {n!} B_n^{(d)} (a) .\label{ber}
\eeq
We use the residue theorem to see that
\beq
\mbox{Res }_{s=z}\zeta_{{\mathcal B}} (s,a) = \frac{(-1)^{d+z} }{(z-1)! (d-z)!}
                    B_{d-z}^{(d)} (a) ,\label{barn}
\eeq
for $z=1,...,d$. The needed leading poles are
\beq
\mbox{Res }_{s=d}\zeta_{{\mathcal B}} (s,a) &=&  \frac 1 {(d-1)!} ,\nn\\
 \mbox{Res }_{s=d-1} \zeta_{{\mathcal B}} (s,a) &=&
  \frac{d-2a}{2 (d-2)!} , \nn\\
 \mbox{Res }_{s=d-2} \zeta _{{\mathcal B}} (s,a) &=&
 \frac {12 a^2 -d-12 a d +3d^2}
{24 (d-3)!} ,\nn\\
\mbox{Res }_{s=d-3} \zeta _{{\mathcal B}} (s,a) &=&
\frac{-8a^3+12a^2d+2a-6ad^2-d^2+d^3}{48 (d-4)!}.\nn
\eeq
We may now determine the residues of
$\zeta (s)$. At $s=\frac {m-3} 2=\frac{d-2} 2$ we find
\beq
\mbox{Res }_{s=\frac {m-3} 2} A_{-1} (s) &=& -\frac {d_s} 6 
     \,\frac{m-2}{2^m \Gamma ((m-1)/2) \Gamma ((m-3)/2)} , \nn\\
\mbox{Res }_{s=\frac {m-3} 2} A_{0} (s) &=& \frac{d_s}{96 \Gamma (m-4)} , \nn\\
\mbox{Res }_{s=\frac {m-3} 2} A_1  (s) &=&\frac {d_s} 6
 \,\frac{(5m-13)}{2^m \Gamma ((m-1)/2) \Gamma ((m-3)/2)} , \nn\\
\mbox{Res }_{s=\frac {m-3} 2} A_2(s) &=&-\frac {d_s} {256} 
     \,\frac{(m-3)^2 (5m-9)}{\Gamma (m-1)} . \nn
\eeq
To get these representations, the `doubling formula' 
$\frac{\Gamma (z)} {\Gamma (2z)} = \frac{\sqrt{2\pi} 2^{1/2-2z} }
{\Gamma (z+1/2)}\nn$ for the $\Gamma$ function 
and its functional relation 
$\Gamma (z+1) = z\Gamma (z)$ has been used. Summing up, using 
again the given properties of the $\Gamma$-functions and (\ref{ex1}) for 
the heat-kernel coefficient $a_3$, we find
\beq
a_3 &=& 2^{-5-m} (m-1) d_s \frac{8 (4m-11) \Gamma ( m/2) +(17-7m)
 \Gamma (1/2) \Gamma ((m+1)/2)} {3\Gamma (m/2) \Gamma ((m+1)/2) } \nn\\
&=& (4\pi ) ^{-(m-1)/2} \int_{S^{m-1}} 
 \operatorname{Tr}\left[
      \frac{(4m-11) (m-1) \Gamma (m/2)}{48\Gamma (1/2) \Gamma ((m+1)/2)} 
      +\frac{(17-7m)(m-1)}{384} \right] \nn\\
   &=& (4\pi ) ^{-(m-1)/2} \int_{S^{m-1}}
 \operatorname{Tr}\left[
 d_{16} (m-1) +d_{17} (m-1)^2 \right] .\nn
\eeq
Form here, equation (10a) is immediate. 

To get equations (10b) and (10c) we need to introduce a smearing function. 
For our purposes a smearing function 
of the form $F(r) = f_0 +f_1 r^2 +f_2 r ^4$ is suitable. 
We note that the radial 
normalization constant is given by $C= 1/J_{p+1} (\mu )$. We denote 
the normalized Bessel function by
$$\bar J _p (\mu r
) := J_p (\mu r ) / J_{p+1} (\mu ).$$
Instead of the
zeta function we consider now the smeared analogue:
\beq
\zeta (F;s) = \sum_\lambda \int_{B^m} F(x) \varphi ^* (x) \varphi (x) \frac 1
{\lambda^{2s} }. \label{zetasmear}
\eeq
Since $F$ depends only on the normal variable, the integral in equation
(\ref{zetasmear}) over the sphere $S^{m-1}$ behaves as in the case $F=1$ so that
\begin{eqnarray}
&& \zeta (F;s) = 4\sum_{n=0} ^\infty d_n (m) \int_{{\mathcal C}} \frac {dk}{2\pi
 i}
k^{-2s}\\
&& \qquad\qquad\cdot\int_0^1 dr F(r) r (\bar J^2_{p+1} (kr) + \bar J_p ^2 (kr)
)
\frac \partial {\partial k} \ln J_p (k) .\label{kk10}
\end{eqnarray}
The radial integrals may be computed using Schafheitlin's reduction 
formula \cite{watson}:
\beq
\lefteqn{
(j+2) \int^z dx x^{j+2} J_\nu ^2 (x) = (j+1) \left\{ \nu^2 -\frac{(j+1)^2} 4
\right\} \int^z dx x^j J_\nu^2 (x) }\nn\\
& & +\frac 1 2 \left[z^{j+1} \left\{ z J_\nu ' (z) -\frac 1 2 (j+1) J_\nu (z) 
\right\}^2+z^{j+1} \left\{ z^2 -\nu^2 +\frac 1 4 (j +1)^2 \right\} 
       J_\nu^2 (z)\right] . \nn
\eeq
For the case at hand, using $J_p (\mu ) =0$, we find the radial integrals
\beq
 \int_0^1 dr\,\,r^3 \left[ \bar J ^2_p (\mu r) + \bar J _{p+1} ^2 (\mu r )
\right] &=&  \frac{2p^2 +3p +1} {3\mu^2} +\frac 1 3 , \nn\\  
 \int_0^1 dr\,\,r^5\left[ \bar J ^2_p (\mu r) + \bar J _{p+1} ^2 (\mu r )
\right] &=&  \frac{8p^4 +20p^3 -20 p -8}{15 \mu^4} +\frac
{4p^2+10p +4}{15\mu^2} +\frac 1 5 .\nn
\eeq
Substituting these into (\ref{kk10}) the contour integral representations
for $\zeta (r^2; s)$ and $\zeta (r^4;s)$ are easily given. The resulting
expressions are evaluated using equation (\ref{ex2}); simple substitutions
suffice to evaluate all relevant terms analogous to 
(\ref{ex11})---(\ref{ex13a}).
The factors of $1/\mu^2$ and $1/\mu^4$ are absorbed by using $s+1$ and $s+2$
instead of $s$ in equations (\ref{ex11})---(\ref{ex13a}). The powers 
of $p$ lower the argument of the base zeta function by $2$, by $3/2$, 
by $1$, by $1/2$ and by $0$.
It is now a
straightforward matter to compute:
\beq
A_{-1} (r^2;s) &=&
\frac 1 {4\Gamma( \frac{1}{2})} 
\frac {\Gamma (s-\frac 1 2)}{\Gamma (s+1)} \zeta
_{S^d} (s-\frac 1 2) 
\left[\frac 1 3 +\frac 2 3 \frac{s-\frac 1 2}{s+1} \right] \nn\\
& & +\frac 1 {4\Gamma( \frac{1}{2})} 
\frac{\Gamma (s+\frac 1 2)}{\Gamma (s+2)}\left[
\zeta_{S^d} (s) +\frac 1 3 \zeta_{S^d} (s+\frac 1 2) \right] ,\nn\\
A_0 (r^2;s) &=&
-\frac 1 4 \zeta _{S^d} (s) -\frac 1 4 \zeta _{S^d} (s+\frac 1 2)
-\frac 1 {12} \zeta_{S^d} (s+1) , \nn\\
A_1 (r^2; s) &=& 
-\frac 2 {3\Gamma (s+1)} \zeta_{S^d} (s+\frac 1 2)\left[
\frac 1 {8\Gamma( \frac{1}{2})} 
\Gamma (s+\frac3 2) -\frac 5 {12 \Gamma( \frac{1}{2}) }
\Gamma (s+\frac 5 2) \right]  \nn\\
& & -\frac 1 {3\Gamma (s)} \zeta_{S^d} (s+\frac 1 2)\left[
\frac 1 {8\Gamma( \frac{1}{2})} 
\Gamma (s+\frac 1 2) -\frac 5 {12 \Gamma( \frac{1}{2}) }
\Gamma (s+\frac 3 2) \right]\nn\\
& &-\frac 1 {\Gamma (s+1)} \zeta_{S^d} (s+1) 
         \left[ \frac 1 {8\Gamma (1/2)} \Gamma (s+3/2) -\frac 5 
        {12 \Gamma (1/2)} \Gamma (s+5/2)\right] + ...\,\, ,\nn\\
A_2 (r^2; s) &=& -\frac 2 {3\Gamma (s+1) } \zeta_{S^d} (s+1)
    \left[ \frac 1 {16}\Gamma (s+2) -\frac 3 8 \Gamma (s+3) +\frac 5 {32} 
    \Gamma (s+4) \right] \nn\\
 & &+\frac 1 {3\Gamma (s)} \zeta_{S^d} (s+1)
    \left[ \frac 1 {16}\Gamma (s+1) -\frac 3 8 \Gamma (s+2) +\frac 5 {32}
    \Gamma (s+3) \right]+...\nn
\eeq
This exemplifies very well the rules of substitution and we spare to write
down the associated terms for $\zeta (r^  4; s)$ explicitly. 

Although lengthy, it is again easy to add up all contributions to find 
$a_3 (F,D,{\cal B})$ for the smearing function given by $F(r) = f_0 +f_1 r^2 
 +f_2 r^4$. We derive equations (10b) and (10c) by identifying the boundary invariants:
\beq
F(1) &=& F\left|_{\partial M} = f_0 +f_1 +f_2 ,\right. \nn\\
F' (1) &=& -F_{;m} \left|_{\partial M}= 2f_1 +4f_2 , \right.\nn\\
F'' (1) &=& F_{;mm}  \left|_{\partial M}= 2f_1 +12f_2 . \right.\ \qed\nn
\eeq

\medbreak\noindent{\bf Proof of (11).} We give the ball $B^2$ the usual metric
$ds_B^2=dr^2+r^2d\theta^2$. Let $N$ be a compact Riemannian manifold without boundary
and let $M=B^2\times N$ with the product metric. The extrinsic curvature is $L_{\theta 
\theta} =1$, $L_{ab} =0$ otherwise. Let $\tilde P$ be the Dirac operator 
on $N$. The Dirac operator $P$ on $M$ reads 
\beq
P = \left( \frac{\partial}{\partial x_m} -\frac 1 {2r} \right) 
 \gamma _{m(m)} +\frac 1 r 
 \left( \bea {cc}
 0  & \sqrt{-1} \gamma_{\theta (m-1)} \\
 -\sqrt{-1} \gamma_{\theta (m-1)} & 0 
  \eea \right) \partial _\theta + 
 \left( \bea {cc}
 0  & \sqrt{-1} \tilde P \\ 
 -\sqrt{-1} \tilde P & 0 
 \eea \right) . \label{n1}
\eeq
Write the eigenfunction $\varphi$ of $P$, $P\varphi = \mu 
\varphi$, in the form
$\varphi = {\psi_1 \choose \psi_2}$. Let ${\cal Z}_n$ be an eigenfunction of 
$\tilde P$. An ansatz of the form $\psi_1 = f(r) e^{i(m+1/2)\theta} {\cal Z}
_n$ is not possible because $\gamma_{\theta (m-1)}$ and $\tilde P$ anticommute.
A simultaneous set of eigenfunctions 
of $\partial_\theta$ and $\tilde P$ thus does not exist. However, $\gamma
_{\theta (m-1)}$ plays the role of '$\gamma^5$' for the $\gamma$-matrices 
on $N$. Therefore, define ${\cal Z}_n^\pm$ to be the upper and lower chirality
parts of ${\cal Z}_n$,
\beq
\czp := \frac 1 {\sqrt 2} \left( 1\pm \sqrt{-1} \gamma_{\theta (m-1)}\right) 
      \cz .\nn
\eeq
Consequently
\beq
\tilde P \czp = \lambda_n \cz^\mp \mbox{ and }\tilde P^2 \czp 
      = \lambda_n^2 \czp , \nn
\eeq
and $\psi_1 = f(r) e^{i(m+1/2)\theta} \czp$ might be chosen. A full set 
of eigenfunctions is then found to read
\beq
\varphi_1^{(\pm )} &=& e^{i(m+1/2) \theta} 
     \left( \bea{c} 
       J_{m+1} ( \rel r) \cz^+ \\
   \mp \frac i \mu \rel J_{m+1} (\rel r ) \cz ^- \mp 
      \frac{i\lambda_n} \mu J_m (\rel r) \cz ^+ 
      \eea 
\right), \label{sol1}\\
\varphi_2^{(\pm)}  &=& e^{i(m+1/2) \theta}
\left( \bea{c} 
       J_m (\rel r ) \cz ^- \\
       \pm \frac i k \rel J_{m+1} (\rel r) \cz  ^- \mp \frac{i\lambda_n} \mu
     J_m (\rel r ) \cz ^+ 
      \eea \right) . \label{sol2} 
\eeq
We need to impose spectral boundary conditions. We choose $\theta = 1/2$. 
the boundary operator reads 
\beq
A= \left( \bea {cc}
      \gamma_{\theta (m-1)} & 0 \\
   0 & -\gamma_{\theta (m-1)} 
   \eea \right) \partial _\theta + 
     \left( \bea {cc}
   -\tilde P & 0 \\
   0 & \tilde P \eea \right) \nn
\eeq
and we need the projection on its non-negative spectrum. Obviously one chooses 
the ansatz $\alpha = {\alpha_1 \choose \alpha_2}$ as eigenspinor of $A$
and gets the 
equations
\beq
\gamma_{\theta (m-1)} \partial_\theta \alpha_1 -\tilde P \alpha_1 &=& 
         E_t \alpha_1 , \nn\\
-\gamma_{\theta (m-1)} \partial_\theta \alpha_2 +\tilde P \alpha_2 &=&
         E_t \alpha_2 , \label{aps1}
\eeq
Define 
$
b_\pm = \frac{m+1/2 \pm \sqrt{\lambda_n^2 +(m+1/2)^2}} \lambda .\nn
$
Expand $\alpha_1$ and $\alpha_2$ in terms of $\czp$. Then eigenfunctions are given by:
\beq
\alpha_1^{(\mp)} &=& e^{i(m+1/2)\theta} (b_\pm \cz^+ +\cz^-) \text{ and }
\alpha_2 ^{(\mp)} = e^{i(m+1/2)\theta} (b_\mp \cz^+ +\cz^-) ,\text{ where }\\\nn
A\alpha^\mp & = &\mp \sqrt{\lambda_n^2 +(m+1/2)^2} \alpha^\mp . \nn
\eeq
Imposing spectral boundary conditions so means that the projection on 
all eigenfunctions $\alpha^+$ has to vanish. Boundary conditions can not 
be imposed on $\varphi_1^{(\pm)}$ and $\varphi_2 ^{(\pm )}$, but instead on  
suitable linear combinations. Define
\beq
a_\mp = \frac{\lambda_n\mp \mu} {\rel} ,\nn
\eeq
and impose boundary conditions on $\varphi_1 
+ a_\mp \varphi_2$. This gives the 
conditions, using $b_-b_+ = -1$,
\beq
J_m (\rel ) +\frac{b_-}{a_-} J_{m+1} (\rel ) &=& 0 , \nn\\
J_m (\rel ) +\frac{b_-}{a_+} J_{m+1} (\rel ) &=& 0 . \nn
\eeq
With $a_- a_+=-1$ this can be combined to read
\beq
J_m^2 (\rel ) -\frac{2\lambda_n b_-}{\rel} J_m (\rel ) J_{m+1} (\rel ) 
   -b_-^2 J_{m+1} (\rel ) =0 .\nn
\eeq
So the starting point for the zeta function with smearing function 
$F=1$ is 
\beq
\zeta (s) &=& \sum_{m=-\infty}^\infty \sum_n \int_{{\cal C}} \frac{dk}{2\pi i} 
           (k^2+\lambda_n^2)^{-s} 
           \times 
   \frac{\partial}{\partial k} \ln \left\{ 
        J_m^2 (k) -\frac{2\lambda_n b_- } k J_m (k) J_{m+1} (k) 
      -b_-^2 J_{m+1}^2 (k) \right\} . \nn
\eeq
Using for $l\in \nats$, $J_{-l} (k) = (-1)^l J_l (k)$ and shifting the 
contour to the imaginary axis we find
\beq
\zeta (s) &=& \frac{ 2\sin (\pi s)} \pi 
     \sum_{m=0}^\infty \sum_n \int_{|\lambda_n| } dk \,\, 
        (k^2-\lambda_n^2)^{-s} \times \nn\\
      & &\frac{\partial}{\partial k} \ln \left\{k^{-2m} \left[
        I_m^2 (k) -\frac{2\lambda_n b_- } k I_m (k) I_{m+1} (k)
      +b_-^2 I_{m+1}^2 (k)\right] \right\} . \nn
\eeq
The role of the base zeta function will here be played by the zeta function 
associated with $A^2$. We thus define (actually, this is only $1/2$ 
the zeta function because the sum over $m$ runs from $m=0$ only instead of 
$m=-\infty$) 
\beq
\zeta_A (s) = \sum_{m=0}^\infty \sum_n \left[(m+1/2)^2 
+\lambda_n^2\right]^{-s}
\label{zetaa1}
\eeq
and will need furthermore
\beq
\zeta_A ^l (s) = \sum_{m=0}^\infty \sum_n \frac{(m+1/2)^l}{[(m+1/2)^2 +
   \lambda_n^2 ]^{s}} . \label{zetaa2}
\eeq
This suggests, that a suitable expansion parameter is $\nu = m+1/2$.
We define 
\beq
\delta = \frac \nu {\sqrt{\nu^2 +\lambda_n^2}}\nn
\eeq
and have the following relations,
\beq
\delta = \frac{1-b_-^2}{1+b_-^2} , \quad 
\frac{\delta -1} \delta = \frac {\lambda_n} \nu b_- , \quad
b_-^2 = \frac{1-\delta}{1+\delta} , \quad 
1+b_-^2 = \frac 2 {1+\delta}.\nn
\eeq
In addition, the zeta function associated with the spectrum $\lambda_n$ 
of the manifold $N$ will naturally appear in the calculations,
\beq
\zeta_N (s) = \sum_n (\lambda_n^2) ^{-s}.\nn
\eeq
After a lengthy calculation using the expansion  
(\ref{ex4}) we find for the relevant expression the following asymptotic
expansion for $\nu\to\infty$: 
\beq
& &\ln \left\{ z^{-2\nu +1} \left[ 
      I^2_{\nu -1/2} (z\nu)+b_-^2 I^2_{\nu +1/2} (z\nu) -\frac{2\lambda_n b_-
             }{\nu z} I_{\nu -1/2} (z\nu ) I_{\nu +1/2} (z\nu ) \right]
      \right\} \sim \nn\\
& &\ln \left\{ z^{-2\nu} \frac{e^{2\nu\eta}}{2\pi \nu} (1+b_-^2) 
      \left( 1+t\frac{\sqrt{\lambda^2 +\nu^2}}\nu \right) \right\}+\frac 1 \nu M_1 (t) +\frac
1 {\nu^2} M_2 (t) +{\cal O} (1/\nu^3) .\nn
\eeq
The polynomials are 
\beq
M_1 (t) &=& \frac \delta 2 t^2 -\frac 5 {12} t^3 , \nn\\
M_2 (t) &=& \frac 1 2 \frac{\delta^2}{\delta+t} t^3 
     +\frac 1 8 \frac \delta {\delta+t} t^4 
  -\frac 1 8 \frac{\delta^3}{\delta+t} t^4-\frac 1 2 \frac 1 {\delta+t} t^5 -\frac 5 8
\frac{\delta^2} 
      {\delta+t} t^5 
      +\frac 5 8 \frac 1 {\delta+t} t^7 . \nn
\eeq
In analogy to the treatment in the proof of (10), this suggests the definitions
\beq
A_{-1} (s) &=& \frac{2\sin (\pi s)} \pi \sum_{m=0}^\infty \sum_n 
   \int_{|\lambda_n|/\nu}^\infty dz \,\, (z^2\nu^2-\lambda_n^2)^{-s} 
           \frac{\partial}{\partial z} 
                \ln \left( z^ {-2\nu} e^{2\nu\eta}\right) , \nn\\
A_0 (s)&=& \frac{2\sin (\pi s)} \pi \sum_{m=0}^\infty \sum_n
   \int_{|\lambda_n|/\nu}^\infty dz \,\, (z^2\nu^2-\lambda_n^2)^{-s}
           \frac{\partial}{\partial z}
                \ln \left(1+t\frac{\sqrt{\lambda_n^2 +\nu^2}}\nu\right),\nn\\
A_q (s) &=& \frac{2\sin (\pi s)} \pi \sum_{m=0}^\infty \sum_n
   \int_{|\lambda_n|/\nu}^\infty dz \,\, (z^2\nu^2-\lambda_n^2)^{-s}
           \frac{\partial}{\partial z}
             \frac{M_q (t)}{\nu^q} .\nn
\eeq
We use (\ref{hyper}) to see 
\beq
A_{-1} (s) = -\frac 2 {\sqrt{\pi} \Gamma (s)} \int_{{\cal C}} \frac{dt}{2\pi i}
    \,\, \frac{\Gamma (s-1/2 +t) \Gamma (-t)}{t-1/2}\zeta_H (-2t; 1/2) 
                  \zeta _N (s+t-1/2) , \label{amecon}
\eeq
where the contour lies to the left of $\Re t = -1/2$. If we denote the 
heat-kernel coefficients of $\tilde P^2$ on $N$ as $a_j ^{(N)}$, we have the 
relations \cite{BGb}:
\beq
\Gamma ((m-2)/2) \mbox{ Res }_{s=(m-2)/2} \zeta_N (s) &=& a_0^{(N)} = 
          (4\pi)^{-(m-2)/2} \int_N  \trace \,\, 1, \nn\\
\Gamma ((m-4/2) \mbox{ Res }_{s=(m-4)/2} \zeta_N (s) &=& a_1^{(N)} =
          (4\pi)^{-(m-2)/2} \left( -\frac 1 {12}\right) 
          \int_N  \trace R(N) . \nn
\eeq
For later use, in the same way we define $a_j^{(S^1\times N)}$ associated 
with $A^2$.
Using $\zeta_A (s)$ instead of $\zeta_N (s)$ in the above equations,
the results with obvious replacements remain valid.

Shifting the contour in (\ref{amecon}) to the left  
we pick up the poles of $A_{-1} (s)$. 
To provide checks of the calculation, we also present the residues
to the 
right of $s=(m-3)/2$.
E.g.~we find that 
\beq
\Gamma (m/2) \mbox{ Res }_{s=m/2} A_{-1} (s) &=& \frac 1 2 a_0^{(N)} ,\nn\\
\Gamma ((m-1)/2)\mbox{ Res }_{s=(m-1)/2} A_{-1}  (s) &=&0,\nn\\
\Gamma ((m-2)/2)\mbox{ Res }_{s=(m-2/2} A_{-1} (s) &=&\frac 1 2 a_1^{(N)}
             -\frac 1 {12} a_0^{(N)},\nn\\
\Gamma ((m-3)/2)\mbox{ Res }_{s=(m-3)/2} A_{-1}  (s) &=&0.\nn
\eeq
We continue with $A_0 (s)$. It may be casted into the form
\beq
A_0 (s) = -\frac 1 {\Gamma (s)} \zeta_{S^1\times N} (s) \sum_{k=0}^\infty 
         (-1)^k \frac{\Gamma (s+(k+1)/2)}{\Gamma ((k+3)/2)} .\nn
\eeq
At the values of $s$ needed the $k$-sum can be given in closed form and 
one finds
\beq
\Gamma ((m-1)/2)\mbox{ Res }_{s=(m-1)/2} A_0(s)&=& -a_0^{(S^1\times N)} 
        \left\{1-\frac{\Gamma (m/2)} {\Gamma (1/2) \Gamma ((m+1)/2)} \right\},
          \nn\\
\Gamma ((m-2)/2)\mbox{ Res }_{s=(m-2)/2} A_0(s)&=&0,\nn\\
\Gamma ((m-3)/2)\mbox{ Res }_{s=(m-3)/2} A_0(s)&=& -a_1^{(S^1\times N)}
       \left\{1-\frac{\Gamma (m/2-1)} {\Gamma (1/2) \Gamma ((m-1)/2)} \right\}.
       \nn
\eeq
Similarly, $A_1 (s)$ and $A_2 (s)$ can be represented in terms of $\zeta_A ^l 
(s)$, equation (\ref{zetaa2}). The relevant residues of $\zeta_A^l (s)$ 
can be determined from $\zeta_A (s)$ by a suitable scaling of the circle 
$S^1$. One has 
\beq
\sum_{m=0}^\infty \sum_n \frac{(\nu^2)^l}{(\lambda_n^2 +\nu^2 )^{s+l+1}} 
          &=& (-1)^l \frac{\Gamma (s+1)}{\Gamma (s+l+1)} \times\left( \frac d {db} \right)^l \sum_{m=0}^\infty \sum_n
                 (\lambda_n^2 +\nu^2 b)^{-s-1} \left|_{b=1}. \right.\nn
\eeq
The residues of the right hand side can be obtained from $a_j^{(S^1 \times N)}$.
E.g.~
\beq
\mbox{Res }_{s=(m-3)/2} \sum_{m=0}^\infty \sum_n (\lambda_n^2 +\nu^2 b)^{-s-1}
           = \frac 1 {\Gamma ((m-1)/2)} \frac{a_0^{(S^1 \times N)} }{\sqrt b}
      .\nn
\eeq
It follows
\beq
\mbox{Res }_{s=(m-3)/2} \sum_{m=0}^\infty \sum_n 
 \frac{(\nu^2)^l}{(\lambda_n^2 +\nu^2 )^{s+l+1}}
= \frac{\Gamma (l+1/2)}{\Gamma (1/2) \Gamma (m/2 +l-1/2)} a_0^{(S^1 \times N)}.
\nn
\eeq
This, and a similar equation for $s=(m-2)/2$, allows one to find the remaining 
contributions to the leading pole:
\beq
& &\textstyle\Gamma (\frac{m-2}{2}) \mbox{ Res }_{s=(m-2)/2} A_1 (s) = 
     \frac 1 3 \left( 1-\frac 3 4 \frac{\Gamma (1/2) \Gamma (m/2)}
         {\Gamma ((m+1)/2) } \right) (4\pi )^{-m/2} 
               \int_{\partial M}  \trace 1,\nn\\
& &\Gamma ((m-3)/2) \mbox{ Res }_{s=(m-3)/2} A_2(s) =\nn\\
& &      \left(-\frac{3m-4}{16\Gamma (1/2) (m^2-1)} 
         \frac{\Gamma (m/2)} {\Gamma ((m+1)/2)} + 
           \frac{m^2+8m-17}{128 (m^2-1)} \right)(4\pi)^{-(m-1)/2} 
              \int_{\partial M}  \trace 1 .\nn
\eeq
Putting things together, we can use $a_0$, $a_1$ and $a_2$ as a check 
of the calculation. The value we compute for
$d_{12}$ agrees with our previous calculation. Finally, we complete the proof of assertion
(11) of Lemma 4. \qed
\bigbreak
{\bf Acknowledgement:} We would like to thank Stuart Dowker for 
very interesting and helpful discussions on the subject.
PG has been supported by the NSF (USA) and MPI
(Leipzig).
KK has been supported by the EPSRC under Grant No GR/M45726.
\bigbreak

\end{document}